\documentclass[fleqn,11pt]{article}
\usepackage{hyperref}
\usepackage{amsmath, amssymb, booktabs, multirow, graphicx, footnote, caption, color}

\usepackage{tabularx}

\usepackage{geometry}
\usepackage{algorithm}  
\usepackage{algpseudocode}
\usepackage{booktabs}

\usepackage{booktabs}
\usepackage{adjustbox}
\usepackage{array}
\usepackage{titlesec}
\usepackage{amsfonts}  
\usepackage{graphicx}                    
\usepackage{float, epsfig}
\usepackage{latexsym}
\usepackage{footnote}
\usepackage{amsthm}
\usepackage{url}
\usepackage{subfigure}
\makeatletter
\newcommand{\Rmnum}[1]{\expandafter\@slowromancap\romannumeral #1@}
\makeatother
\begin{document}\sloppy
	
	\title{Forecasting solar power output in Ibadan: A machine learning approach leveraging weather data and system specifications}
	
	\author{\footnote{${}^{3}$ Corresponding author: Email address: caston.sigauke@univen.ac.za, Phone No: +27 15 962 8135 (C Sigauke)}\bf ${}^{1}$Obarotu Peter Urhuerhi, ${}^{2}$Christopher Udomboso and ${}^{3*}$Caston Sigauke \\ \\
		${}^{1}$Center for Petroleum, Energy Economics and Law, University of Ibadan, Ibadan, Nigeria \\
		${}^{2}$Department of Statistics, University of Ibadan, Ibadan, Nigeria \\
		${}^{3}$Department of Mathematical and Computational Sciences, \\ University of Venda, Private Bag X5050, Thohoyandou \\ 0950, Limpopo, South Africa}
	\maketitle

\begin{abstract} This study predicts hourly solar irradiance components, Global Horizontal Irradiance (GHI), Direct Normal Irradiance (DNI), and Diffuse Horizontal Irradiance (DHI) using meteorological data to forecast solar energy output in Ibadan, Nigeria.
The forecasting process follows a two-stage approach: first, clear-sky irradiance values are predicted using weather variables only (e.g., temperature, humidity, wind speed); second, actual (cloudy-sky) irradiance values are forecasted by integrating the predicted clear-sky irradiance with weather variables and cloud type. Historical meteorological data were preprocessed and used to train Random Forest, Convolutional Neural Network (CNN), and Long Short-Term Memory (LSTM) models, with Random Forest demonstrating the best performance. Models were developed for annual and seasonal forecasting, capturing variations between the wet and dry seasons. The annual Random Forest model’s normalised Root Mean Square Error (nRMSE) values were 0.22 for DHI, 0.33 for DNI, and 0.19 for GHI. For seasonal forecasts, wet season nRMSE values were 0.27 for DHI, 0.50 for DNI, and 0.27 for GHI, while dry season nRMSE values were 0.15 for DHI, 0.22 for DNI, and 0.12 for GHI. The predicted actual irradiance values were combined with solar system specifications (e.g., maximum power (Pmax), open-circuit voltage (Voc), short-circuit current (Isc), and AC power (Pac)) using PVLib Python to estimate the final energy output. This methodology provides a cost-effective alternative to pyranometer-based measurements, enhances grid stability for solar energy integration, and supports efficient planning for off-grid and grid-connected photovoltaic systems.\\ 
	
\noindent{\bf Keywords:} Neural Networks, Random Forest, Renewable energy, Solar power forecasting, Weather data. 
\end{abstract}

\section{Introduction} \label{sec:1.0}
\subsection{Background and motivation} \label{sec:1.1}
\begin{figure}[H]
	\centering
	\includegraphics[width=0.8\textwidth]{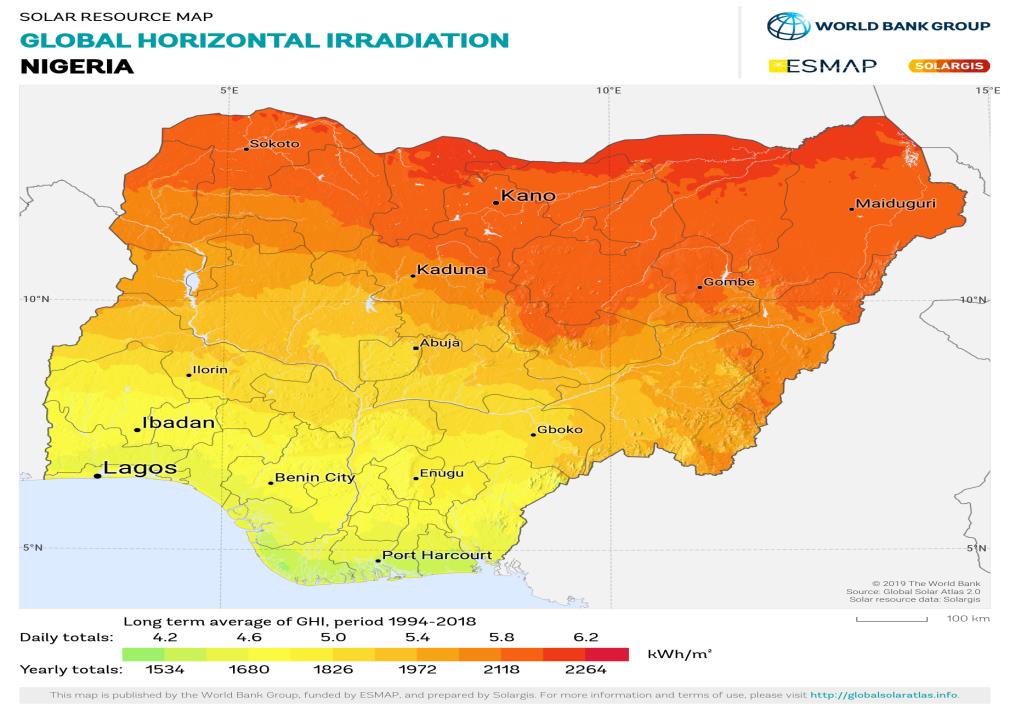}
	\caption{Map of the global horizontal irradiance of Nigeria \cite{Solargis2024}.}
	\label{fig:ghi_map}
\end{figure}

In the context of Nigeria, the energy landscape presents its unique challenges. The country’s electricity supply is unreliable, with frequent power outages and an inadequate grid infrastructure failing to meet rising energy demands \cite{Leahy2019, Ikeanyibe2021}. This situation has led businesses, individuals, and government agencies to explore alternative energy sources, particularly renewable energy. Among these alternatives, solar energy, especially solar photovoltaic (PV) systems, has gained considerable attention as a reliable solution \cite{Olukan2022}. Across the country, solar energy is widely available and offers a reliable, clean power source less prone to outages. As illustrated in Figure \ref{fig:ghi_map}, the Global Horizontal Irradiance (GHI) in Nigeria varies from around 3.5 kWh/m²/day in coastal parts to 7.0 kWh/m²/day in northern locations \cite{Solargis2024, Ohunakin2014}.

The Nigerian government has aggressively promoted renewable energy as a key component of its plan to address the nation’s energy difficulties after realising the potential of solar power. This involves leveraging solar energy through programs like the National Renewable Energy and Energy Efficiency Policy (NREEEP) and the Nigeria Solar Energy Policy (NSEP) \cite{Ozoegwu2021}.
\par
Rural electrification is one of the most urgent problems in Nigeria’s energy industry. Many Nigerians, especially those living in rural regions, still do not have access to power. An estimated 70\% of these rural areas live in energy poverty since they are not linked to the national electricity grid \cite{Iwayemi2008}. The Nigerian government has started several rural electrification initiatives to solve this issue and provide more access to energy in these neglected areas. As of 2015, eight Power Purchase Agreements (PPAs) had indicated interest in establishing solar power projects, according to a report from the Nigerian Bulk Electricity Trading Company (NBET) \cite{Isoken2016}.
\par
Deploying off-grid and mini-grid solutions powered by renewable sources is one of the tactics, with solar energy at the forefront. These off-grid solar solutions offer a practical and scalable approach to bringing electricity to remote and rural communities where extending the national grid may not be economically or logistically feasible. By harnessing solar energy, these rural electrification projects aim to provide reliable, clean, and sustainable power to millions of Nigerians without access to electricity \cite{Babalola2022, Elusakin2014}.
\par 
However, intermittent solar energy presents a problem for off-grid and grid-connected applications. It is challenging to forecast and control the energy supply because solar system output mostly relies on meteorological factors \cite{Alzahrani2017, Antonanzas2016}. This fluctuation emphasises the importance of creating precise forecasting models to estimate solar energy output. Whether for rural electrification initiatives or as a supplement to the national grid, these models are essential for guaranteeing the dependability of the power supply  \cite{Moreno2008, Antonanzas2016, Voyant2017, Wang2019}.
\par
Thus, this study aims to investigate advanced techniques for predicting solar energy production in Nigeria, enhancing the precision and dependability of these projections. By doing this, it hopes to encourage Nigeria’s wider adoption of solar energy, aid in attempts to electrify both urban and rural areas and assist in providing sustainable energy to fulfil the nation’s expanding energy needs.
\par
While there has been research on forecasting Global Horizontal Irradiance (GHI) in Nigeria, there is a need for continued efforts to improve accuracy. This research will focus on predicting GHI, direct normal irradiance (DNI) and diffuse horizontal irradiance (DHI), as both DNI and DHI are crucial for accurately calculating the output energy of solar systems. Given that solar energy generation is highly site-dependent, this study will concentrate on Ibadan, using the Center of Petroleum, Energy Economics and Law (CPEEL), University of Ibadan (Lat 7.4515$^o$N, Lon 3.8899$^o$E) to develop a more precise predictive model tailored to the local environment.

\subsection{Literature Review} \label{sec:1.2}
Machine learning (ML) and artificial neural networks (ANN) have emerged as essential tools in energy modelling, proving their adaptability across renewable and non-renewable energy sectors. In the oil and gas sector, for instance, these techniques have been successfully applied to predict oilfield scale formation, forecast gas production trends, and model crude oil prices \cite{Falode2016, Falode_and_Udomboso2016, Falode_and_Udomboso2021}. These applications highlight the ability of ML and ANN to handle complex datasets and provide accurate predictions, thereby enhancing operational efficiency and economic analysis in energy systems.
\par 
Building on this foundation, the current review focuses on applying traditional statistical learning models and artificial neural networks (ANN) in solar energy research. This section begins by examining significant global contributions to the field. From this worldwide perspective, it explores solar energy research undertaken in Nigeria, tackling the unique challenges and opportunities relevant to this setting. Lastly, it refines its focus to studies centred on Ibadan, Nigeria, which acts as the case study for this research.
\par
David et al. \cite{David2016} draw comparisons between solar time series and financial data by probabilistically forecasting solar irradiance on a global scale using a combination of ARMA and GARCH models. With fewer parameters and faster implementation, the recursive ARMA-GARCH hybrid exhibits accuracy equivalent to machine learning methods like Support Vector Machines and Gaussian Process. The hybrid model’s biggest advantage is its ability to provide real-time forecasts without requiring a training period. However, forecast accuracy decreases with increasing lead time and under variable sky conditions. Sites with stable weather, like Desert Rock, showed better forecast accuracy, while more variable locations, such as Oahu and Fouillole, presented greater challenges in predictability. This study highlights the importance of sky variability in solar forecasting, as more dynamic conditions lead to less accurate predictions.
\par
Voyant et al. \cite{Voyant2012} adopted a hybrid ARMA-ANN model for predicting hourly global radiation. They optimise a Multi-Layer Perceptron (MLP), combining it with an ARMA model. This approach outperforms traditional models, achieving a normalised Root Mean Square Error (nRMSE) of 14.9\% compared to 26.2\% for a naïve persistence model. The hybrid model also improves accuracy over stand-alone ANN models, with an average nRMSE gain of 3.5\%. The study emphasises the importance of stationarity in time series forecasting and suggests that the hybrid model’s performance is robust and does not result from data mining.
\par
Wu et al. \cite{Wu2014} investigated forecasting the short-term photovoltaic (PV) output, which is crucial for solar power system management. They examined variables that impact PV output, such as solar irradiance, module temperature, and cloud cover, using historical data from PV systems in Malaysia and Taiwan. The research produced five forecasting models: ANN, ANFIS, ARIMA, Support Vector Machines (SVM), and a hybrid model that combined these with a Genetic Algorithm (GA). The models’ goal was to forecast PV output one hour in advance, and solar irradiance was found to be a critical input. According to the results, the hybrid model had the maximum accuracy and efficiency.
\par
In the Thar desert region of Rajasthan, India, Chandola et al. \cite{Chandola2020} provide a solar irradiance forecasting model using Long Short-Term Memory (LSTM) networks for 3 hours, 6 hours and 12 hours ahead forecast. Variables including temperature, dew point, wind direction, wind speed, humidity, and GHI, DHI, and DNI are all included in the model. The study uses data from four locations, Jaisalmer, Bikaner, and Jodhpur- from 2010 to 2014. Using RMSE and MAPE, the performance of the LSTM model was assessed. The Bikaner region has the best RMSE of 0.099 and MAPE of 4.54\% for forecasting 3 hours in advance. Jodhpur and Jaisalmer produced the best 6-hour and 24-hour forecasts, with RMSE values of 0.129 and 0.117, respectively. The study suggests that the model can effectively forecast solar radiation in arid regions, aiding in the design of optimised solar energy systems. They concluded that future work could explore hybrid ARIMA-LSTM models for improved accuracy.
\par
Benali et al. \cite{Benali2018} assess the forecasting of hourly solar irradiance components (GHI, DNI, and DHI) at Odeillo, France, a region renowned for its high levels of climatic fluctuation, using Smart Persistence, Artificial Neural Networks (ANN), and Random Forest (RF). The study uses 1-minute resolution data for 1 to 6 hours of time horizons. RF outperforms both ANN and Smart Persistence, with the lowest normalised root mean square error (nRMSE) for GHI (19.65\%- 27.78\%), DNI (34.11\%-49.08\%), and DHI (35.08\%-49.14\%). RF’s accuracy improves with longer forecasting horizons, and forecasting is less reliable during spring and autumn due to increased meteorological variability. The study confirms RF as the most effective method among those tested.
\par
In order to forecast solar irradiance in Canada, Alzahrani et al. \cite{Alzahrani2017} applied a Long Short-Term Memory (LSTM). When the model was tested against Support Vector Regression (SVR) and Forward Neural Networks (FNN), LSTM outperformed. As compared to SVR (0.11) and FNN (0.16), LSTM had the lowest RMSE (0.086) in the results, indicating its greater accuracy. The research demonstrated how well deep learning models—specifically LSTM—predict solar irradiance, with average Mean Bias Error (MBE) being as low as possible.
\par
Gbémou et al. \cite{Gbémou2021} compare how well machine learning models predict Global Horizontal Irradiance (GHI) using data taken every ten minutes in their investigation. Artificial Neural Networks (ANNs), Support Vector Regression (SVR), and Gaussian Process Regression (GPR) are among the models that were tested. The study is conducted over two years (2018–2019), using training data from 2018 and testing data from 2019. With an emphasis on predicting GHI at time horizons from 10 minutes to 4 hours, the research highlights multi-step forward forecasting. The findings demonstrate that machine learning models perform better than the scaled persistence model, especially LSTM and GPR with a rational quadratic kernel. Normalised root mean square error (nRMSE), coverage width-based criterion (CWC), and dynamic mean absolute error (DMAE) was used to compare the performance of the models. With nRMSE values of 0.2016 at 10 minutes and 0.3995 at 4 hours, LSTM outperformed other techniques, particularly when considering longer time horizons.
\par
Ghimire et al. \cite{Ghimire2019} presented the CLSTM. This hybrid deep learning model combines Long Short-Term Memory Networks (LSTM) and Convolutional Neural Networks (CNN) for half-hourly forecasting of Global Horizontal Irradiance (GHI) in Alice Springs, Australia. The model relies on CNN for data feature extraction and LSTM for prediction. The research shows that CLSTM performs better than stand-alone models such as CNN, LSTM, RNN, and DNN, with a Mean Absolute Percentage Error (MAPE) of 4.84\%. Over 70\% of predictions are within \(\pm10 \, \text{Wm}^{-2}\), and the hybrid model has a low Relative Root Mean Square Error (\(\approx 1.515\%\)), significantly outperforming the other models.  
\par	
In estimating daily solar radiation in Northwest Nigeria, specifically focusing on Kano, Kaduna, and Katsina, Aliyu et al. \cite{Alivu2020} proposed an Artificial Neural Network (ANN) model. The study aimed to improve the accuracy of daily solar radiation forecasts, which is crucial for solar energy system optimisation. The Nigerian Meteorological Agency (NIMET) provided data spanning 21 years (1994-2015), of which 70\% was used for training and 30\% for testing. The root mean square error (RMSE) and coefficient of determination (R²) were used to assess the ANN model. The R² was found to be 0.78, and the RMSE was found to be 0.47 for training and 0.48 for testing.
\par 
The research by Ismail and Aliyu (2023) focuses on using a Long Short-Term Memory (LSTM) deep learning model to predict daily solar radiation. The study area includes three states in Northwest Nigeria- Kano, Kaduna, and Katsina, where the researchers used 21 years of solar radiation data from the Nigerian Meteorological Agency. Initially, in millijoules per meter squared per day, the data was converted to kilowatt-hours per meter squared per day (kWh/m²/day) for analysis. The LSTM model demonstrated strong predictive capabilities, with a coefficient of determination (R²) of 0.79 for the training dataset and 0.78 for the testing dataset. The model also produced Root Mean Square Errors (RMSE) of 0.46 and 0.47 for training and testing datasets. These results indicate that the LSTM model is effective for daily solar radiation forecasting in the region.
\par 
Adeyemi-Kayode et al. \cite{Adeyemi2022} investigate solar energy characteristics in Nigeria using MATLAB, focusing on diffuse horizontal irradiance (DHI), global horizontal irradiance (GHI), and direct normal irradiance (DNI). The study encompasses data from eleven locations across Nigeria, including Kano, Kaduna, Yola, Port-Harcourt, Eko, Ikeja, and Ibadan. The data from Solcast covers the period from December 30, 2014, to December 31, 2020. The input parameters used in the study are hours, days, months, years, DNI, DHI, and air temperature, while the GHI value is used as the output parameter. The study uses a hybrid Artificial Neural Networks (ANN) and Particle Swarm Optimisation (PSO) model to forecast solar irradiance. The model's performance metrics include a normalised Root Mean Square Error (nRMSE) ranging from 0.7813\% to 13.8948\%, Mean Squared Error (MSE), and Mean Absolute Percentage Error (MAPE). The optimal PSO parameters were C1 = 1.0 and C2 = 2.5, with a maximum of 1200 iterations. Results indicate that northern locations like Kano, Kaduna, and Yola have the highest GHI values compared to southern regions.
\par 
In Uyo, Nigeria, Ntekim and Uppin \cite{Ntekim2024} looked into the application of artificial neural networks (ANNs) for accurate solar energy forecast. By examining the mathematical links between important variables, such as solar radiation, panel size, efficiency, and performance ratio, the main goal was to forecast the best solar power output. The research utilised a 366-day dataset of hourly solar radiation data that included components of direct, diffuse, and reflected radiation from the European Commission for 2020. The components were summed to generate a single hourly radiation data, which was subsequently used to compute hourly power output using equation \eqref{eq1}. These hourly power values were averaged to derive the daily power output, serving as the model's output, while the input consisted of the hourly radiation data.

\begin{equation}
	E = A \cdot r \cdot H \cdot PR,
	\label{eq1}
\end{equation}
where \( A \) is the panel area (\( m^2 \)), \( r \) is the panel efficiency (taken as 22.8\%), \( H \) is the hourly radiation (\( W/m^2 \)) and \( PR \) is the performance ratio and coefficient of losses (taken as 0.75).
\par
The ANN model was trained using 80\% of the dataset, with the remaining 20\% reserved for testing. The model achieves a Mean Absolute Error (MAE) of 0.03 and a Root Mean Squared Error (RMSE) of 0.03 for a panel area of 0.68 \( m^2 \). For a larger panel area of 1.0068 \( m^2 \), the MAE was 0.04 and RMSE was 0.05. The study also found that peak solar power generation occurred around 1 p.m. daily, with January recording the highest average solar power output.
\par
In order to estimate daily GHI in Ibadan, Ayodele et al. \cite{Ayodele2019} created a hybrid k-means and support vector regression (SVR) algorithm. Seasonal data were clustered using K-means, and solar radiation predictions for different sky conditions were made using SVR models. The model was tested using six years of meteorological data (2010–2015) from Ibadan and then used to predict daily global solar radiation for future years (2016–2017). The training features included rainfall, evaporation, wind speed, minimum and maximum relative humidity, and minimum and maximum temperature. Data were sourced from the International Institute of Tropical Agriculture (IITA), Ibadan. Performance was evaluated using RMSE, RRMSE, MAPE, and R². For 2016, the model achieved an R² of 0.9816, RMSE of 0.119 kWh/m²/day, RRMSE of 2.6515\%, and MAPE of 1.7928\%. For 2017, the values were R² of 0.9842, RMSE of 0.121 kWh/m²/day, RRMSE of 2.7498\%, and MAPE of 1.795\%.

\subsection{Contribution and research highlights}
There are considerable gaps in existing solar forecasting research, particularly in the tropical regions like West Africa, where climate-specific research remains limited. Secondly, most models only account for meteorological inputs without considering vital system parameters such as panel efficiency and installation details, which limits their practical applicability in solar energy system planning and optimisation.
\par
\begin{figure}[H]
	\centering
	\includegraphics[width=1.0\textwidth]{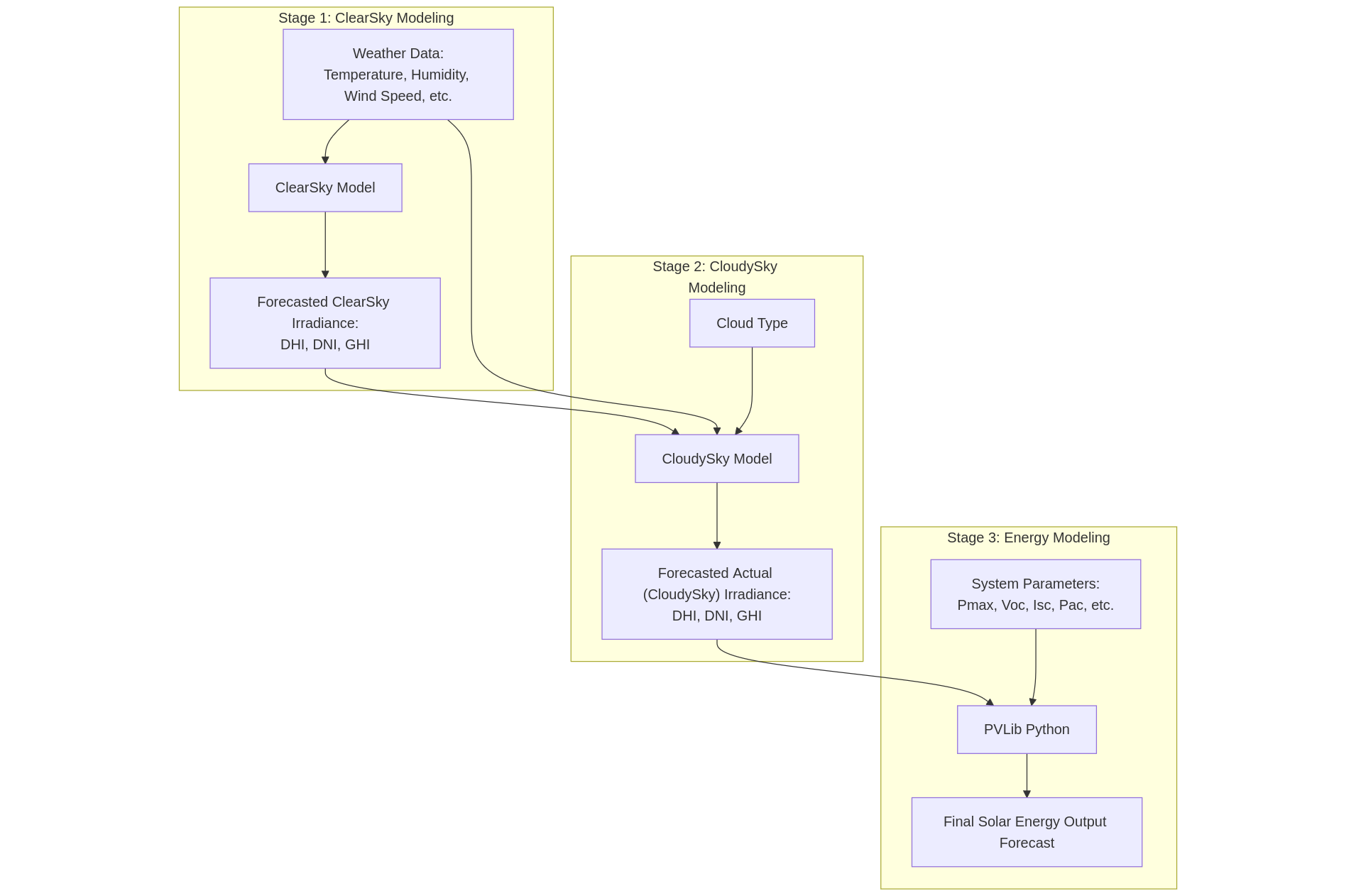}
	\caption{Model Framework for Solar Irradiance and Energy Forecasting.}
	\label{fig:framework}
\end{figure}

 
This study develops a two-step machine learning framework for predicting solar irradiance components (GHI, DNI, DHI) and energy yield in Ibadan, Nigeria, from meteorological inputs and PV system simulation, with Random Forest outperforming CNN and LSTM models. The key highlights of the study are:
\begin{itemize}
	\item  The study uses a two-stage prediction method. First, clear-sky irradiance is predicted using weather variables, followed by cloud type and weather data to predict actual (cloudy-sky) irradiance.
	\item Comparison between machine learning algorithms was made. Random Forest performed the best with an nRMSE of 0.19 (GHI), 0.22 (DHI), and 0.33 (DNI), performing better than CNN and LSTM.
	\item The dry season predictions were more precise (nRMSE: 0.12–0.22) than wet season (nRMSE: 0.27–0.50) because of cloud cover difficulties.
	The calculated irradiance parameters were merged with solar system parameters (like Pmax, Voc) through PVLib Python to calculate the final energy generation. The model framework diagram is presented in Figure \ref{fig:framework}
	\item The study provides an economical alternative to pyranometers, enhances grid stability, and allows for off-grid and grid-connected solar system planning.
\end{itemize}

The rest of the paper is organised as follows. A detailed discussion of the methodology used in this study is given in Section \ref{sec:2.0}. The empirical results and their discussion are provided in Section \ref{sec:3.0}, while Section \ref{sec:4.0} concludes and provides areas for future research.

\section{Methodology} \label{sec:2.0}

The methodology adopted is outlined in the following sections, detailing the Data Collection and Scope processes, Exploratory Data Analysis (EDA), Data Preprocessing, Forecasting Models, Evaluation Metrics and PV-LIB Python modelling. These sections describe the analytical methods and procedures employed to achieve the objectives of this study.

\subsection{Data collection and scope}
The dataset for this research was obtained from the National Solar Radiation Data Base (NSRDB) \cite{Sengupta2018} for the Center of Petroleum, Energy Economics and Law (CPEEL), University of Ibadan (Lat 7.4515°N, Lon 3.8899°E), spanning from 2005 to 2022, collected at hourly intervals. The description of the variables (columns) of the datasets is presented in Table \ref{tab1}.

\begin{table}[H]  
	\centering  
	\small % Reduce font size  
	\caption{Variable Definitions.}  
	\label{tab1}  
	\begin{tabularx}{\textwidth}{lX} % Use tabularx for adjustable column widths  
		\toprule  
		\textbf{Column Name} & \textbf{Description} \\
		\midrule  
		Year & The year of the recorded data. \\
		Month & The month of the recorded data. \\
		Day & The day of the recorded data. \\
		Hour & The hour of the recorded data. \\
		Solar Zenith Angle & The angle between the sun’s rays and the vertical direction, affecting the intensity of solar radiation. \\
		Surface Albedo & The reflectivity of the Earth’s surface at a given location. \\
		Precipitable Water & The total amount of water vapour in a column of the atmosphere. \\
		Clearsky DHI & Diffuse Horizontal Irradiance under clear-sky conditions. \\
		Clearsky DNI & Direct Normal Irradiance under clear-sky conditions. \\
		Clearsky GHI & Global Horizontal Irradiance under clear-sky conditions. \\
		Cloud Type & Classification of Cloud Cover Affecting Solar Radiation. \\
		Dew Point & The temperature at which air becomes saturated with moisture. \\
		Relative Humidity & The moisture percentage in the air relative to the maximum it can hold at that temperature. \\
		Pressure & The atmospheric pressure at the surface level. \\
		DHI & Diffuse Horizontal Irradiance—solar radiation scattered from the atmosphere on a horizontal surface. \\
		DNI & Direct Normal Irradiance—solar radiation received directly from the sun. \\
		Fill Flag & Indicator for whether missing data was filled. \\
		GHI & Global Horizontal Irradiance—the total solar radiation received on a horizontal surface. \\
		Temperature & Ambient temperature at the location. \\
		Wind Direction & The direction from which the wind is blowing. \\
		Wind Speed & The speed of wind at the location. \\
		\bottomrule  
	\end{tabularx}  
\end{table}

\subsection{Exploratory data analysis}
Exploratory Data Analysis (EDA) section focuses on understanding the dataset for modelling. First, we will analyse the descriptive statistics to summarise key characteristics like averages, medians, and ranges. Missing data and outliers will be identified to ensure they do not distort the results. We will use visual tools like box plots to uncover relationships between weather variables and solar radiation components. Correlation analysis will follow to assess the strength of these relationships. Lastly, the data will be divided into wet and dry seasons, and similar analyses will be repeated for each, allowing us to capture seasonal patterns. These steps will lay the groundwork for the modelling and evaluation processes.

\subsection{Data preprocessing}
The data preprocessing stage is a crucial step in machine learning and modelling, which aids in improving data quality and ensures it is ready for analysis. It involves cleaning, transforming, and structuring raw data to make it consistent and usable for building models. By addressing issues like missing values, scaling, and irrelevant features, preprocessing helps enhance model accuracy and reliability.

\subsubsection{Data cleaning}
In this data preprocessing phase, we address missing values, outliers, and irrelevant data to ensure the dataset is well-prepared for analysis. Specifically, we remove nighttime (inutile) data, as solar irradiance is non-existent during these hours, and we remove data with missing cloud properties for cloudy-sky modelling. Additionally, we adjusted the dataset by shifting timestamps forward by 30 minutes (e.g. 06:30 becomes 07:00). This adjustment ensures the data aligns with an hourly basis for prediction.

\subsubsection{Training variables}
To train our models effectively, we identify specific features (input variables) and targets (output variables) tailored to the clear-sky and cloudy-sky scenarios. The clear-sky model’s predictors include weather and temporal features such as Month, Day, Hour, precipitation water, Dew Point, Relative Humidity, Pressure, Temperature, Wind Speed, and Wind Direction. The target variables for this model are the clear-sky irradiance components: Clearsky GHI, Clearsky DNI, and Clearsky DHI.
\par
In the cloudy-sky model, additional features account for dynamic atmospheric conditions, notably the Clear-sky irradiance components and Cloud Type. At the same time, the target variables are the actual observed irradiance components: GHI, DNI, and DHI. This selection of variables ensures that each model is appropriately equipped to predict the irradiance components under varying weather conditions.

\subsubsection{Data splitting}
To have a robust model evaluation, the dataset for the year 2022 is first separated from the modelling process. This data is reserved as a validation dataset and will not be included in the training or testing phases. It will evaluate the model’s performance by comparing predicted values against actual observed values, providing an independent assessment of its predictive capabilities. The remaining dataset, excluding 2022, is then split into two subsets: 80\% for training and 20\% for testing. This ensures the model learns patterns from the most available data while being evaluated on unseen data during testing.

\subsubsection{Data standardisation}
Standardisation will be applied to both the features and target variables to improve consistency and optimise the performance of the machine learning models. The features were standardised using the StandardScaler (equation \eqref{eq2}), which ensures the data has a mean of zero and unit variance. This scaling method is especially useful for models like neural networks, where gradient descent optimisation benefits from features on similar scales, allowing the model to converge faster. This approach effectively handles features with varying magnitudes, ensuring no single feature disproportionately influences the model \cite{Hastie2009}.

\begin{equation}
	z = \frac{x - \mu}{\sigma},
	\label{eq2}
\end{equation}

where \( x \) is the input value, \( \mu \) denotes the mean of the dataset, \( \sigma \) is the standard deviation of the dataset and \( z \) is the standardised value. 
\par
For the target variables, the MinMaxScaler (equation \eqref{eq3}) was employed to scale values to a range between 0 and 1. This aligns well with activation functions such as sigmoid or tanh, whose outputs fall within (0, 1) and (-1, 1), respectively \cite{Bouraya2024}. This work utilises the Rectified Linear Unit (ReLU) activation function, which outputs values in the range of $[0, \infty)]$. Using MinMaxScaler for the target variables also preserves their interpretability, as the scaling maintains the original data’s relative proportions and physical bounds \cite{Raju2020}. These considerations were essential to ensure both computational efficiency and meaningful model outputs.

\begin{equation}
	X' = \frac{X - X_{min}}{X_{max} - X_{min}},
	\label{eq3}
\end{equation}
where \( X' \) is the scaled value, \( X \) represents the input value, \( X_{min} \) is the minimum value of the dataset and 
\( X_{max} \) is the maximum value of the dataset. 

\subsection{Forecasting models}
This paper employs three different models, Random Forest (RF), Convolutional Neural Network (CNN), and Long Short-Term Memory (LSTM), to estimate solar radiation components using a machine learning approach. This section provides a brief review of these models’ mechanics.

\subsubsection{Random Forest}
Since its introduction by Breiman \cite{Breiman2001}, the Random Forest (RF) method has gained widespread recognition for its effectiveness in classification and regression applications. Using bagging (bootstrap aggregating) to provide an extra layer of accuracy and resilience expands on the idea of decision trees. By creating multiple randomised decision trees and aggregating their predictions, random forests increase the stability and predictive accuracy of individual decision trees (equation \eqref{eq4}), especially in situations where the number of input features greatly exceeds the number of observations \cite{Benali2018}.

\begin{equation}
	\hat{y} = \frac{1}{T} \sum_{t=1}^{T} \hat{y}_t(x),
	\label{eq4}
\end{equation}
where \( \hat{y}_t \) is the regression output, \( T \) is the total number of trees in the forest and \( \hat{y}_t(x) \) is the prediction of the \( t \)-th tree for input \( x \).
\par 	
Each tree in the Random Forests starts by splitting the feature space (X) into smaller, more homogeneous subgroups using binary splits. The CART (Classification and Regression Trees) criterion often guides this process, optimising the homogeneity at each step. At the leaves, predictions are based on the mean of the grouped data. Cross-validation is used to prune the trees to prevent overfitting. Breiman’s innovation adds randomisation by selecting a random subset of predictors for each split, enhancing generalisation and reducing overfitting. The final prediction is obtained by averaging results from multiple trees, offering improved accuracy and scalability \cite{Biau2016}.
\par
\begin{figure}[H]
	\centering
	\includegraphics[width=0.8\textwidth]{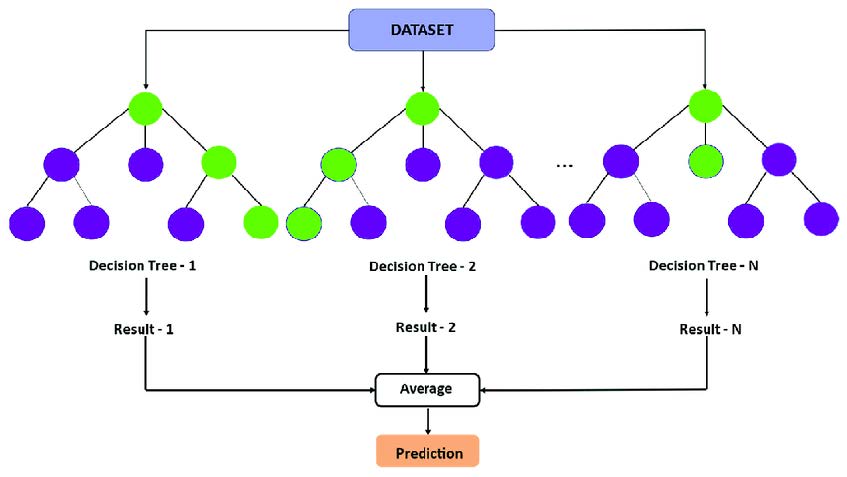}
	\caption{Algorithm for making predictions using Random Forest model \cite{Segovia2023}.}
	\label{fig:random_forest}
\end{figure}

 
The algorithm for making predictions using the Random Forest model is given in Figure \ref{fig:random_forest}.

\subsubsection{Convolutional Neural Network}
Convolutional Neural Networks (CNNs) are a class of Deep learning models known to have demonstrated remarkable efficacy in tackling intricate problems such as image classification, natural language processing, and time series forecasting. Since LeCun et al. \cite{Lecun1998} originally presented CNNs, they have become widely used because of their capability to identify spatial hierarchies of features from unprocessed input data automatically.
\par
\begin{figure}[H]
	\centering
	\includegraphics[width=0.8\textwidth]{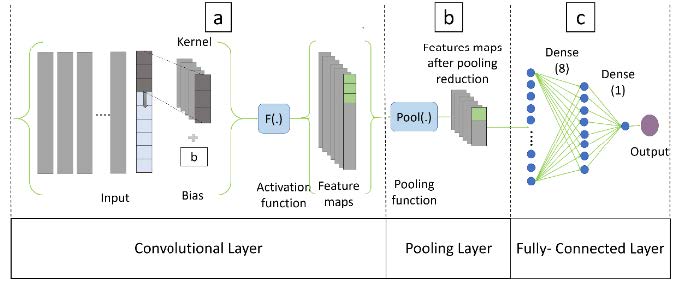}
	\caption{1D-CNN general architecture \cite{Vakitbilir2021}.}
	\label{fig:cnn_architecture}
\end{figure}

Convolutional layers are the first layer of a CNN (Figure \ref{fig:cnn_architecture}), where filters are applied to extract pertinent characteristics from the input data. As it moves over the input space, the filter’s convolution process enables the network to identify local relationships and patterns in the data \cite{Goodfellow2016}. Activation functions like ReLU (Rectified Linear Unit) are added after the convolutional layers to add non-linearity to the model \cite{Albawi2017}.
\par 
There are several types of CNNs, the most popular ones being 1D and 2D CNNs. 1D-CNN is applied to numerical data, whereas 2D-CNN is mainly utilised for text and graphics. While the layers of the two models are comparable, the convolutional layer is where they differ most. Whereas in 1D-CNN, the sliding takes place in only one dimension, in 2D-CNN, kernels slide across the input data in two dimensions \cite{Vakitbilir2021}.
\par
This paper employed a 1D-CNN to predict solar irradiance using weather features such as temperature, humidity, and wind speed. The data was reshaped into a 3D format to enable the convolution operation along the time dimension. CNNs are effective for spatial data (e.g., images) and time series data \cite{Krizhevsky2012}.
\par
The final layers of the CNN consist of fully connected (dense) layers, which aggregate the features learned in previous layers to make a prediction. These dense layers are followed by an output layer that provides the forecasted values for the target variables— DNI, DHI, and GHI. CNNs offer key advantages such as parameter sharing and local receptive fields, making them highly scalable for large datasets \cite{Goodfellow2016}. Their ability to capture local patterns, combined with the depth of the model, makes them a powerful tool for forecasting tasks, especially in solar energy prediction, where patterns in meteorological data can be complex and interdependent. Figure \ref{fig:cnn_architecture} shows the 1D-CNN general structure.

\subsubsection{Long Short Term Memory}
\begin{figure}[H]
	\centering
	\includegraphics[width=0.8\textwidth]{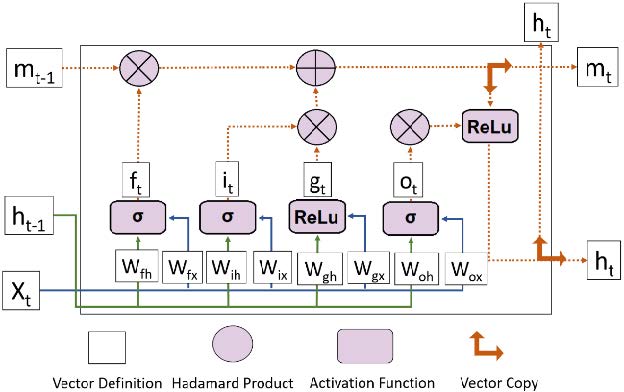}
	\caption{Sample structure LSTM unit \cite{Vakitbilir2021}}
	\label{fig:lstm_unit}
\end{figure}

Introduced by Hochreiter and Schmidhuber \cite{Hochreiter1997}, Long Short Term Memory (LSTM) networks are a variant of Recurrent Neural Networks (RNNs) intended to address the vanishing gradient issue and capture long-range dependencies in time-series data. LSTMs, in contrast to conventional RNNs, employ memory cells to store data for extended periods, enabling the network to “remember” important patterns. The input, output, and forget gates in every cell, as seen in Figure \ref{fig:lstm_unit}, regulate the information flow and allow the network to select which data to store, update, or discard. Gers et al. later added an extra forget gate in 1999 \cite{Hochreiter1997, Gers2000}. LSTMs are widely used for tasks involving sequential data like weather prediction and energy forecasting. The formulas for the forget gate, $f_t$, input gate, $i_t$, intermediate state $g_t$, and output gate $o_t$ are presented in equations \eqref{eq5} – \eqref{eq8}, respectively.

\begin{equation}
	f_t = \sigma(W_{fx}X_t + W_{fh}h_{t-1} + b_f), 
	\label{eq5}
\end{equation}

\begin{equation}
	i_t = \sigma(W_{ix}X_t + W_{ih}h_{t-1} + b_i),
	\label{eq6}
\end{equation}

\begin{equation}
	g_t = \sigma(W_{gx}X_t + W_{gh}h_{t-1} + b_g), 
	\label{eq7}
\end{equation}

\begin{equation}
	o_t = \sigma(W_{ox}X_t + W_{oh}h_{t-1} + b_o),
	\label{eq8}
\end{equation}

where $\sigma$ is the non-linear activation function (sigmoid function), $W_x$ and $W_h$ represent the weight matrices, $X_t$ represents the input of the current time-step, $h_{t-1}$ is the output of the previous time-step, and $b$ is the bias of the relevant gates \cite{Vakitbilir2021, David2023}.
The sample structure of the LSTM unit is shown in Figure \ref{fig:lstm_unit}.

Table \ref{tab:ml_strengths_weaknesses} summarises the strengths and weaknesses of RF, CNN, and LSTM.

\begin{table}[H] 
	\small
	\centering  
	\caption{Strengths and weaknesses of RF, CNN, and LSTM.}  
	\begin{tabular}{|>{\centering\arraybackslash}m{3cm}|>{\arraybackslash}m{5cm}|>{\arraybackslash}m{5cm}|}  
		\hline  
		\textbf{Algorithm} & \textbf{Strengths} & \textbf{Weaknesses} \\
		\hline  
		Random Forest (RF) &   
		\begin{itemize} \setlength{\itemsep}{0pt} \setlength{\topsep}{0pt}
			\item Handles high-dimensional data effectively.  
			\item Robust against overfitting.  
			\item Provides estimates of feature importance.  
			\item Performs well with mixed data types.  
			\item No need for data normalisation or scaling.  
		\end{itemize} &   
		\begin{itemize} \setlength{\itemsep}{0pt} \setlength{\topsep}{0pt}
			\item Less interpretable compared to simpler models.  
			\item Slower inference due to ensemble of trees.  
			\item May perform poorly on imbalanced datasets.  
			\item Sensitive to noisy input features.  
			\item Requires substantial memory and processing time on large datasets.  
		\end{itemize} \\
		\hline  
		Convolutional Neural Network (CNN) &   
		\begin{itemize} \setlength{\itemsep}{0pt} \setlength{\topsep}{0pt}
			\item Highly effective for image and spatial data analysis.  
			\item Learns relevant features automatically.  
			\item Scales well to large datasets.  
			\item Tolerant to input noise and distortions.  
			\item Capable of modelling complex patterns.  
		\end{itemize} &   
		\begin{itemize} \setlength{\itemsep}{0pt} \setlength{\topsep}{0pt}
			\item Requires large labelled datasets.  
			\item Computationally intensive.  
			\item Training can be time-consuming.  
			\item Requires expert tuning of architecture and parameters.  
			\item Risk of overfitting without regularisation.  
		\end{itemize} \\
		\hline  
		Long Short-Term Memory (LSTM) &   
		\begin{itemize} \setlength{\itemsep}{0pt} \setlength{\topsep}{0pt}
			\item Suitable for time-series and sequential data.  
			\item Capable of learning long-term dependencies.  
			\item Handles input variation and missing values well.  
			\item Effective for modelling dynamic temporal patterns.  
		\end{itemize} &   
		\begin{itemize} \setlength{\itemsep}{0pt} \setlength{\topsep}{0pt}
			\item Needs substantial training data for good performance.  
			\item More complex than traditional models.  
			\item Training is computationally expensive due to sequential updates.  
			\item Susceptible to overfitting if not regularised.  
			\item Hyperparameter tuning can be difficult and time-consuming.  
		\end{itemize} \\
		\hline  
	\end{tabular}  
	\label{tab:ml_strengths_weaknesses}  
\end{table}

Applying machine learning models in solar radiation prediction can identify complex relationships and patterns in the data and improve prediction accuracy and reliability, which is vital for solar energy management and applications.

%\subsection{Benchmark model: Generalised additive - tensor product interactions model}
%Let \( y_t \) be GHI at time $t$, i.e. \( t = 1\cdots n \). The generalised additive model with pairwise tensor product interactions of autocorrelated errors is then given as Equations \eqref{eq:tensor1} and \eqref{eq:tensor2}.  
%\begin{equation} 
%	y_t = \beta_{0} t + \sum_{i=1}^{p} s_i (x_{ti}) + \sum_{k=1}^{K} \sum_{j=1}^{J} \tau_{jk} s_j (x_{tj}) s_k (x_{tk}) + \varepsilon_t  
%	\label{eq:tensor1}  
%\end{equation} 
%
%\begin{equation}  
%	\Phi(B)\Phi(B)\varepsilon_t = \theta(B)\Theta(B) v_t  
%	\label{eq:tensor2}  
%\end{equation} 
%where \( \Phi(B) \) is the nonseasonal autoregressive operator, \( \theta(B) \) is the nonseasonal moving average operator, and the corresponding seasonal autoregressive and seasonal moving operators are \( \Phi(B) \) and \( \Theta(B) \) respectively; \( v_t \) denotes a white noise series. By expressing Equation \eqref{eq:tensor1} in terms of \( \varepsilon_t \) and substituting in Equation \eqref{eq:tensor2}, Equation \eqref{eq:tensor3} is produced.  
%\begin{equation}  
%	\Phi(B)\Phi(B)\left[y_t - \left\{ \beta_{0} t + \sum_{i=1}^{p} s_i (x_{ti}) + \sum_{k=1}^{K} \sum_{j=1}^{J} \tau_{jk} s_j (x_{tj}) s_k (x_{tk}) \right\} \right] = \theta(B)\Theta(B)v_t  
%	\label{eq:tensor3}
%\end{equation} 
%
%This study only considered cases where \( j \neq k \), i.e. where quadratic interactions are excluded. The total number of variables, i.e., main effects and interaction variables, is: \( p + \binom{p}{2} = p + \frac{p(p-1)}{2} = \frac{p(p+1)}{2}. \)  
%

\subsection{Evaluation Metrics}
There are no universally accepted standards for model evaluation for solar radiation forecasting, but some metrics are more commonly used in the literature. Among these are RMSE (Root Mean Squared Error), MAE (Mean Absolute Error), nRMSE (normalised RMSE), and ${R}^{2}$ (Coefficient of Determination). When RMSE and MAE values are close, it typically suggests that the model has only small deviations from the actual data, indicating more consistent performance \cite{Gensler2016}.
\par 
While MAPE (Mean Absolute Percentage Error) is also frequently used, it has a significant drawback: it becomes unstable when the actual value $y(i)$ is near zero and is undefined when $y(i) = 0$ \cite{Voyant2017}. Therefore, MASE (Mean Absolute Scaled Error) will be used instead of MAPE. The following sections define the four metrics used in the evaluation.

\subsubsection{Root mean squared error}
Root Mean Squared Error (RMSE) is a commonly used metric for assessing the accuracy of a model by measuring the average magnitude of the error between the predicted and actual values. It is calculated by taking the square root of the average squared differences between predicted and actual values. A lower RMSE indicates better model performance.

\begin{equation}
	RMSE = \sqrt{\frac{1}{n} \sum_{i=1}^{n} (y(i) - \hat{y}(i))^2}
	\label{eq9}
\end{equation}

\subsubsection{Mean absolute error}
Mean absolute error (MAE) is another common metric used to measure the accuracy of a forecasting model. Unlike RMSE, MAE measures the average of the absolute differences between predicted and actual values. MAE does not penalise larger errors as strongly as RMSE, making it a more balanced metric when small and large errors are of equal concern.

\begin{equation}
	MAE = \frac{1}{n} \sum_{i=1}^{n} |y(i) - \hat{y}(i)| 
	\label{eq10}
\end{equation}

\subsubsection{Normalized RMSE}
Normalised RMSE (nRMSE) is often preferred over RMSE because it provides a normalised value that allows easier model comparison. It is obtained by dividing RMSE by the mean of the actual (test) data. This metric enables a more meaningful assessment of model performance, especially when dealing with datasets of different scales.

\begin{equation}
	nRMSE = \frac{RMSE}{\bar{y}(i)} 
	\label{eq11}
\end{equation}

Mohammadi et al. \cite{Mohammadi2015} provided performance thresholds based on nRMSE values, which are given in Table \ref{tab2}.

\begin{table}[H]
	\centering
	\caption{Performance thresholds based on nRMSE values.}
	\begin{tabular}{|c|c|}
		\hline
		\textbf{nRMSE Value} & \textbf{Model Performance} \\
		\hline
		$< 0.10$ & Excellent \\
		$0.10 - 0.20$ & Good \\
		$0.20 - 0.30$ & Fair \\
		$> 0.30$ & Poor \\
		\hline
	\end{tabular}
	\label{tab2}
\end{table}

\subsubsection{Mean absolute scaled error}
The Mean Absolute Scaled Error (MASE) is a robust and effective measure for evaluating time series forecasting that allows for comparing several forecasting models with a naïve benchmark. Scale-invariance and robustness against zero values and outliers are among its key advantages. However, practitioners must be cautious about the constraints around the need for historical data and assumptions on the nature of data when applying MASE in evaluating forecasting results.

The Mean Absolute Scaled Error (MASE) is the ratio of the forecast’s Mean Absolute Error (MAE) to that of a naïve forecasting method. The naïve method usually takes the actual value in the previous period as the forecast (or takes the last seen value for seasonality forecasts). The definition can be expressed mathematically in equation \eqref{eq:mase}. 

\begin{equation} 
	\text{MASE} = \frac{\text{MAE of forecast}}{\text{MAE of naive forecast}} = \frac{\frac{1}{n} \sum_{t=1}^{n} |y_t - \hat{y}_t|}{\frac{1}{n-1} \sum_{t=2}^{n} |y_t - y_{t-1}|},  
	\label{eq:mase}
\end{equation}

where $y_t$ represents the actual value at time $t$, $\hat{y}_t$ is the forecasted value at time $t$, $n$ is the total number of time steps in the dataset.

%\subsection{Coefficient of determination}
%R² indicates how well the predicted values from the model approximate the actual data. It measures the proportion of variance in the dependent variable that is predictable from the independent variables. An R² value close to 1 indicates a better fit of the model to the data.
%
%\begin{equation}
%	R^2 = 1 - \frac{\sum_{i=1}^{n} (y(i) - \hat{y}(i))^2}{\sum_{i=1}^{n} (y(i) - \bar{y})^2},
%	\label{eq12}
%\end{equation}
%
%where $\hat{y}(i)$ is the predicted value for the $i$-th observation in the dataset, $y(i)$ is the actual value for the $i$-th observation, $\bar{y}$ represents the mean or average of all the actual values in the dataset, and $n$ is the total number of observations or data points in the dataset.

\subsection{PV-LIB Python Modelling}
We will leverage PVLib Python to model solar radiation and predict energy output for photovoltaic systems. The process involves six main steps:

\begin{enumerate}
	\item \textbf{Define Location and System Parameters:} The first step involves specifying the system configuration and the site’s geographical attributes. Using PVLib’s \texttt{Location} class, we will define the site’s latitude, longitude, and altitude. The photovoltaic system parameters are created with the \texttt{pvlib.pvsystem.PVSystem} class, where we specify details about the modules and inverters used. For this study, the Trina Solar TSM-500DE18M (II) module is sourced from the California Energy Commission (CEC) database (version 2023.10.31; see Table \ref{tab:trina_module_parameters}). Similarly, the Canadian Solar CS3Y-500MS (2021) module specifications will be extracted from the datasheet provided by the manufacturer. The extra parameters not usually provided in a PV module’s datasheet will be generated using \texttt{pvlib.ivtools.sdm.fit\_cec\_sam()} (see Table \ref{tab:canadian_parameters}). The inverter selected for the modelling is the Fronius Primo GEN24 3.8 208-240 model, sourced from the CEC database (see Table\ref{tab:fronius_inverter_parameters}). Both modules are rated 500W, and the inverter is rated 3800W. See the appendix for full parameter specifications for the modules and inverter.
	
	\item \textbf{Simulate Solar Position:} The solar position, defined by zenith and azimuth angles, will be calculated using the \texttt{get\_solarposition} method of the \texttt{Location} object. This calculation depends on the geographical location and time. The \texttt{get\_solarposition} function utilises latitude, longitude, altitude, and a timestamp to determine the sun’s position.
	
	\item \textbf{Compute Total Irradiance:} Using the solar position derived in the previous step along with the predicted radiation components, we will calculate the total plane-of-array (POA) irradiance using the \texttt{pvlib.irradiance.get\_total\_irradiance} function. The sky diffuse component of the total POA is calculated using the Hay-Davies sky diffuse model \cite{Hay1980} presented in equation \eqref{eq13}.
	
	\begin{equation}
		I_d = DHI \left(A \cdot R_b + (1 - A) \left(1 + \frac{\cos \beta}{2}\right)\right) 
		\label{eq13}
	\end{equation}
	
	DHI represents the diffuse horizontal irradiance, $A$ is the anisotropy index given as the ratio of the direct normal irradiance to the extraterrestrial irradiation, $R_b$ is the projection ratio, given as the ratio of the cosine of the angle of incidence (AOI) to the cosine of the zenith angle, and $\beta$ represents the tilt angle of the array.
	
	\item \textbf{Estimate Module Cell Temperature:} The Faiman model \cite{Faiman2008} presented in equation \eqref{eq14} will be used to estimate the cell temperature of the PV modules, implemented via the \texttt{pvlib.temperature.faiman()} function. This step incorporates the global POA irradiance obtained from the previous step, the ambient temperature, and wind speed.
	
	\begin{equation}
		T_m = T_a + \frac{G_{POA}}{u_0 + u_1 V} 
		\label{eq14}
	\end{equation}
	
	where $T_m$ is the temperature of the module, $T_a$ is the ambient temperature, $G_{POA}$ is the global plane of array irradiance, and $V$ is the wind speed. $u_0$ is the constant heat transfer component (W/m²K) and $u_1$ is the convective heat transfer component (W/m³sK) \cite{Sandia2024}.
	
	\item \textbf{Energy Modelling:} The \texttt{ModelChain} class in PVLib integrates the location and system parameters. The \texttt{run\_model\_from\_poa} method is used to compute the system’s energy output. This process incorporates the POA irradiance, module cell temperature, (ambient/air) temperature, wind speed, precipitable water, and system losses to predict DC and AC power generation. The output of this modelling includes effective irradiance, cell temperature, and predicted power values.
	
	\item \textbf{Data Presentation:} The predicted energy outputs will be visualised through plots, highlighting the performance of selected solar modules and inverters under varying conditions.
\end{enumerate}

\section{Results and discussion} \label{sec:3.0}

\subsection{Exploratory data analysis}
The summary statistics (Table \ref{tab3}) provide insights into weather variables and solar radiation components. Precipitable water, with a mean of 4.65 mm, ranges from 0.8 mm to 6.9 mm, indicating moderate variability and generally stable atmospheric moisture levels, although seasonal shifts exist. The dew point, averaging 22.81°C, fluctuates between 5°C and 27.3°C, reflecting the humid tropical climate. Relative humidity exhibits an average of 88.44\%, with a standard deviation of 13.87\%. The range extends from 19.28\% to 100\%, suggesting that while the region is predominantly humid, occasional extreme dry spells occur. Temperature varies from 11.3°C to 35.5°C, with an average of 25.11°C, showing typical day-night and seasonal shifts. Each parameter has a uniform count (157,776), suggesting no missing data. Wind patterns exhibit variability, with wind direction covering the full 360° and speeds ranging from 0.1 m/s to 5.6 m/s. While wind direction is unpredictable, the low wind speeds point to generally calm conditions.

\begin{table}[H]
	\small
	\centering
	\caption{Summary Statistics for the entire dataset.}
	\begin{tabular}{|l|c|c|c|c|c|c|c|c|c|}
		\hline
		\textbf{Variable (Unit)} & \textbf{Count} & \textbf{Mean} & \textbf{Std} & \textbf{Min} & \textbf{25\%} & \textbf{50\%} & \textbf{75\%} & \textbf{Max} & \textbf{Range} \\
		\hline
		Precipitable Water (cm) & 157776 & 4.65 & 0.99 & 0.80 & 4.20 & 4.9 & 5.4 & 6.9 & 0.8 - 6.9 \\
		Dew Point (°C) & 157776 & 22.81 & 2.49 & 5.00 & 22.40 & 23.4 & 24.2 & 27.3 & 5.0 - 27.3 \\
		Relative Humidity (\%) & 157776 & 88.44 & 13.87 & 19.28 & 79.95 & 94.4 & 100.0 & 100.0 & 19.28 - 100.0 \\
		Pressure (mbar) & 157776 & 987.37 & 1.92 & 980 & 986 & 987 & 989 & 994. & 980.0 - 994.0 \\
		Temperature (°C) & 157776 & 25.11 & 2.91 & 11.30 & 23.00 & 24.6 & 27.3 & 35.5 & 11.3 - 35.5 \\
		Wind Direction (deg) & 157776 & 198.79 & 59.40 & 1.00 & 190.0 & 210.0 & 229.0 & 360.0 & 1.0 - 360.0 \\
		Wind Speed (m/s) & 157776 & 1.50 & 0.77 & 0.10 & 0.90 & 1.3 & 2.0 & 5.6 & 0.1 - 5.6 \\
		Clearsky DHI (W/m²) & 157776 & 94.45 & 123.89 & 0.00 & 0.00 & 0.0 & 172.0 & 670.0 & 0.0 - 670.0 \\
		Clearsky DNI (W/m²) & 157776 & 237.70 & 299.11 & 0.00 & 0.00 & 0.0 & 524 & 934.0 & 0.0 - 934 \\
		Clearsky GHI (W/m²) & 157776 & 268.96 & 346.26 & 0.00 & 0.00 & 0.0 & 589 & 1021 & 0.0 - 1021 \\
		DHI (W/m²) & 157776 & 117.39 & 161.45 & 0.00 & 0.00 & 0.0 & 220 & 740.0 & 0.0 - 740 \\
		DNI (W/m²) & 157776 & 115.92 & 211.62 & 0.00 & 0.00 & 0.0 & 133 & 921.0 & 0.0 - 921 \\
		GHI (W/m²) & 157776 & 201.81 & 284.82 & 0.00 & 0.00 & 0.0 & 392 & 1020 & 0.0 - 1020 \\
		\hline
	\end{tabular}
	\label{tab3}
\end{table}

Solar radiation components display high variability. Clear-sky DNI, GHI, and DHI have mean values of 237.7 W/m², 268.96 W/m², and 94.45 W/m², respectively. Both DNI and GHI exhibit high standard deviations and ranges—up to 934 W/m² for DNI and 1021 W/m² for GHI—indicating that they are heavily influenced by changing weather conditions, which makes them more difficult to predict. In contrast, DHI has a narrower range and lower variability, making it more stable and easier to model.
\par 
For the actual radiation components, DHI averages 117.39 W/m², while DNI and GHI average 115.92 W/m² and 201.81 W/m², respectively. GHI stands out with the highest standard deviation of 284.82 W/m², followed by DNI at 211.62 W/m², while DHI has the lowest variability with a standard deviation of 161.45 W/m².

\begin{figure}[H]
	\centering
	\includegraphics[width=1.0\textwidth]{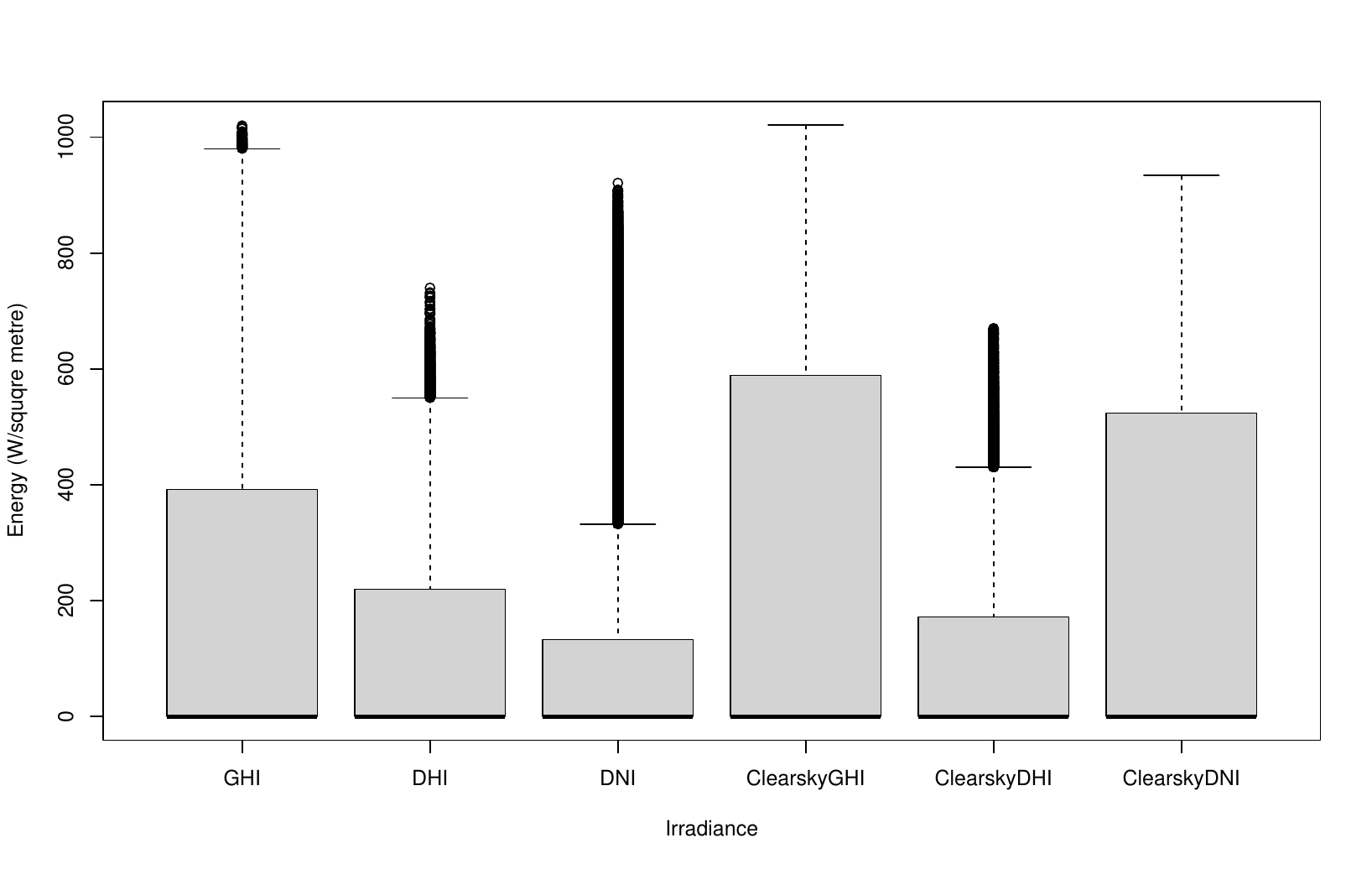}
	\caption{Box plots of irradiance data.}
	\label{fig:GHI_Predictors}
\end{figure}

\begin{figure}[H]
	\centering
	\includegraphics[width=1.0\textwidth]{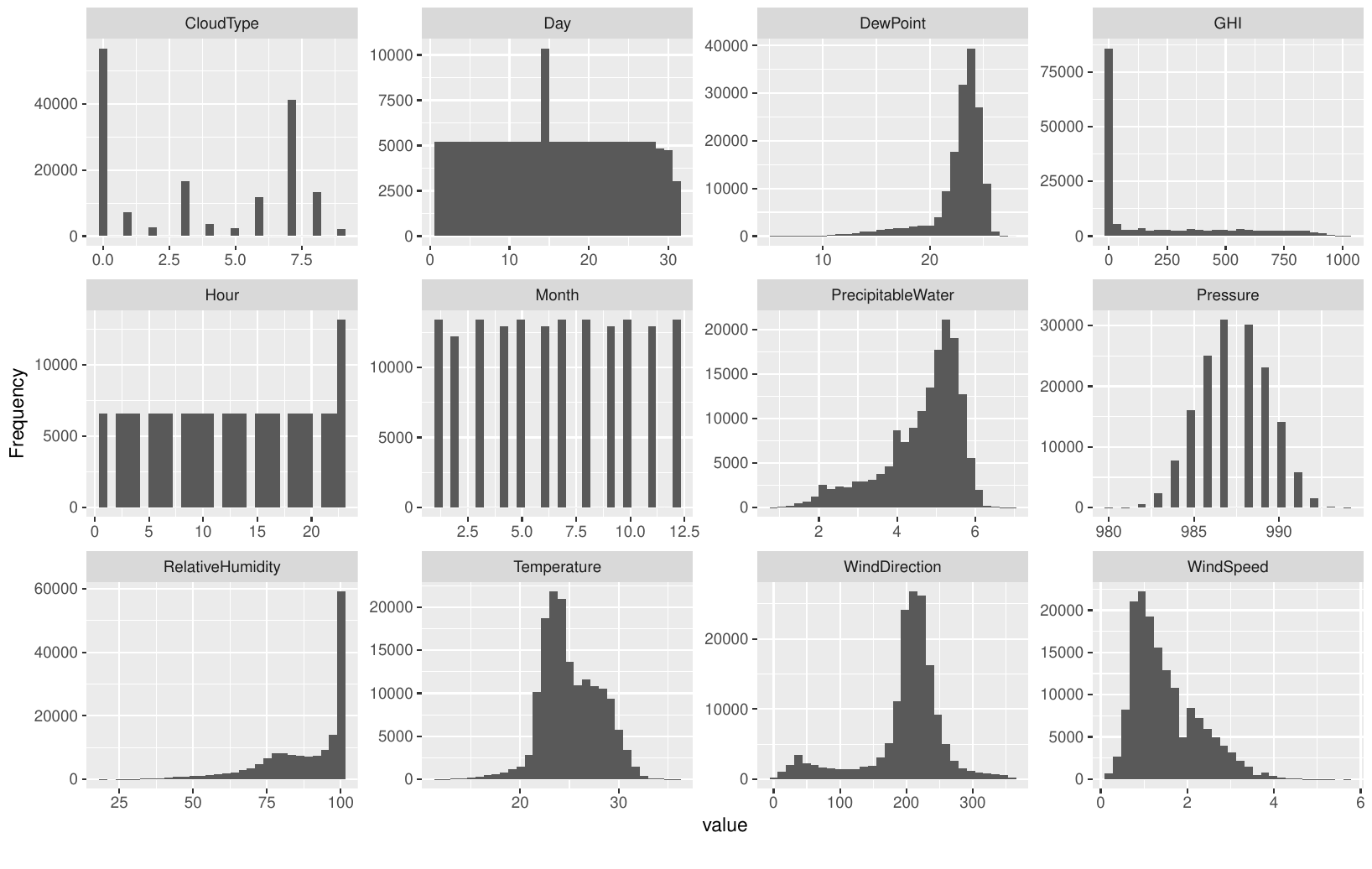}
	\caption{Empirical distributions of variables.}
	\label{fig:dist}
\end{figure}

\begin{figure}[H]
	\centering
	\includegraphics[width=0.8\textwidth]{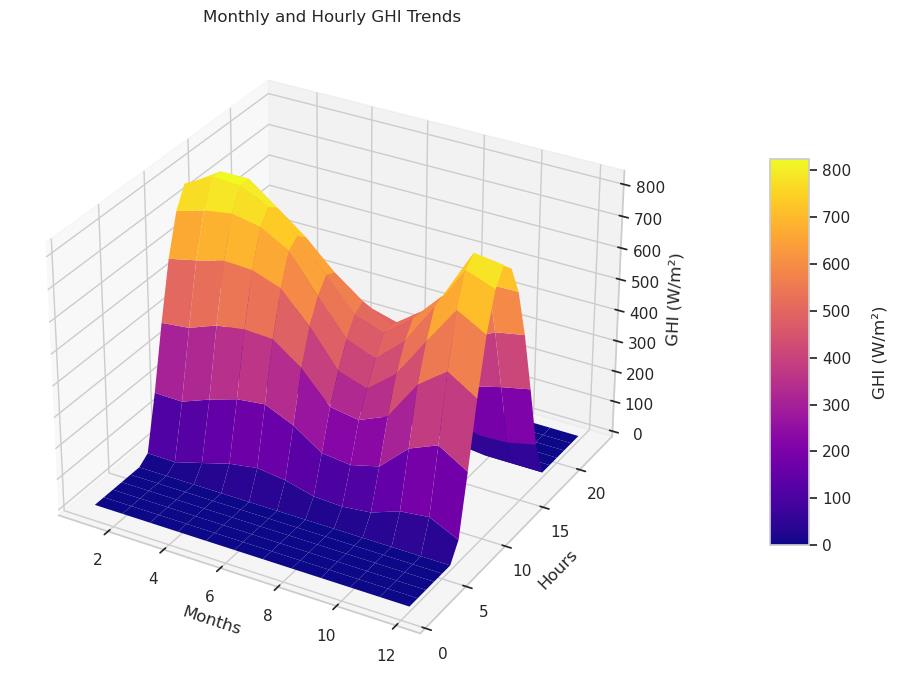}
	\caption{Monthly and hourly solar variation of GHI.}
	\label{fig:ghi_variation}
\end{figure}

Figure \ref{fig:ghi_variation} shows GHI’s seasonal and daily patterns. On the X-axis (months), GHI is highest from January to April and November to December, with a noticeable drop starting in May, marking the start of the wet season, which aligns with \cite{Adelekan2014} division of wet and dry seasons. On the Y-axis (hours), GHI peaks between 11 a.m. and 2 p.m., reflecting the sun’s daily path. This plot highlights the variations in radiation across seasons (wet and dry) and the sun’s daily movement, showing how GHI changes monthly and hourly.

%
%\begin{figure}[H]
%	\centering
%	\includegraphics[width=0.8\textwidth]{fig.t2_hr_month}
%	\caption{Monthly and hourly solar variation of GHI.}
%	\label{fig:ghi_variation1}
%\end{figure}

\begin{figure}[H]
	\centering
	\includegraphics[width=0.8\textwidth]{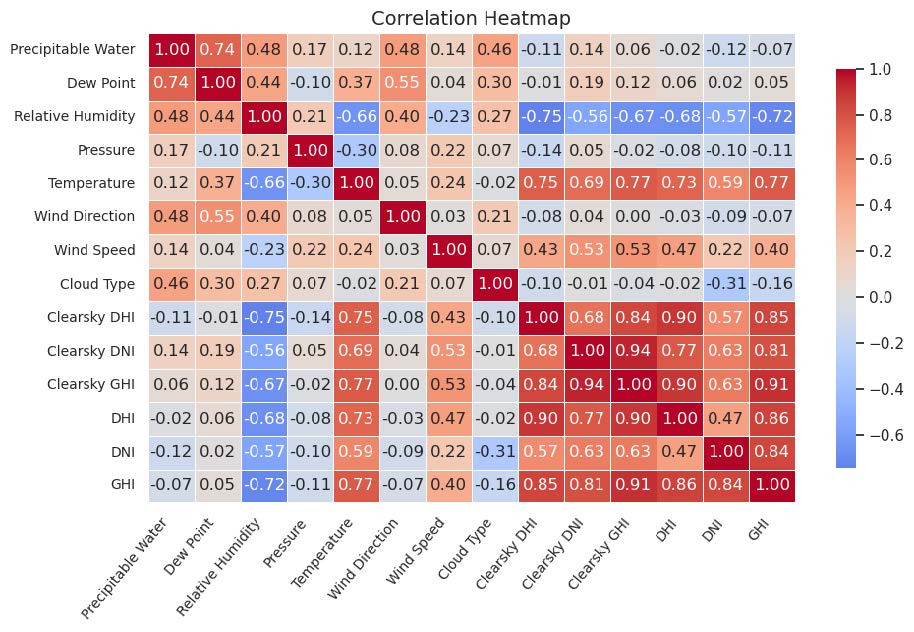}
	\caption{Correlation matrix of weather variables and solar radiation components.}
	\label{fig:correlation_matrix}
\end{figure}

\begin{figure}[H]
	\centering
	\includegraphics[width=0.8\textwidth]{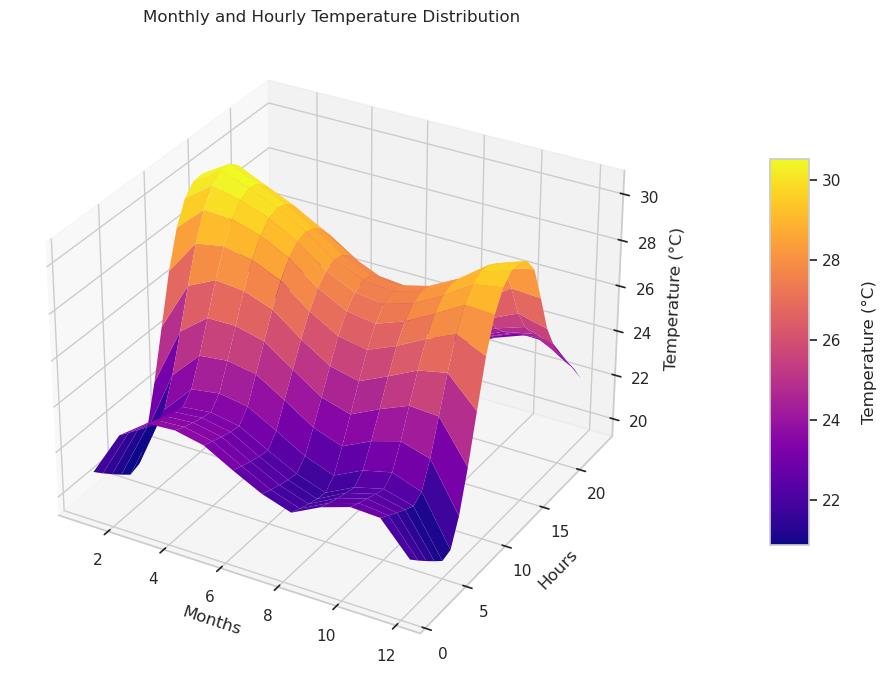}
	\caption{Monthly and hourly temperature variation.}
	\label{fig:temp_variation}
\end{figure}

The correlation plot in Figure \ref{fig:correlation_matrix} highlights the variables strongly correlated with the actual solar radiation components—GHI, DNI, and DHI. Apart from the clear-sky components, which are expectedly highly positively correlated, temperature exhibits the strongest correlation with the solar radiation components, particularly GHI. This relationship is further illustrated in Figure \ref{fig:temp_variation}, where the seasonal and diurnal temperature distribution demonstrates a similar pattern to GHI’s, underscoring their close association. Other weather variables like relative humidity and wind speed also significantly correlate with GHI, DNI, and DHI, but to a lesser degree than temperature.

Figure \ref{fig:temp_variation} shows the monthly and hourly temperature variation.

\subsection{Model evaluation}
To have a robust evaluation, the dataset for the year 2022 was first separated and set aside as a validation dataset. This data was excluded from both the training and testing phases. Instead, it was used to assess how well the model could predict solar radiation components (DHI, DNI, and GHI) by comparing the predicted values with the actual observed values.
\par 
The remaining data, excluding the 2022 dataset, was filtered to remove the nighttime data (7 p.m. – 6 a.m.) and then split into two parts: 80\% for training and 20\% for testing. The DHI, DNI, and GHI predictions and relevant weather variables like temperature, wind speed, and precipitable water were then fed into PVLIB. PVLIB was used to calculate and simulate the corresponding energy output using these inputs. 
\par 
Table \ref{tab4} presents the evaluation results for the annual forecasting with the best results in bold. The irradiance metrics show that the Random Forest (RF) model performed best overall. It recorded the lowest RMSE for DHI (50.6), DNI (77.02), and GHI (77.55) and the lowest MAE for DHI (27.52), DNI (35.57), and GHI (41.36). For nRMSE, RF tied with CNN and LSTM for DHI (0.22), matched LSTM for DNI (0.33), and achieved the lowest score for GHI (0.19). RF also had the lowest MASE for DHI (0.15), DNI (0.13) and GHI (0.13), confirming its reliability for actual irradiance forecasting.

\begin{table}[H]
	\centering
	\caption{Model comparison for annual forecasting.}
	\begin{tabular}{|l|c|c|c|c|c|c|c|}
		\hline
		\textbf{Metric} & \textbf{Model} & \textbf{Clearsky DHI} & \textbf{Clearsky DNI} & \textbf{Clearsky GHI} & \textbf{DHI} & \textbf{DNI} & \textbf{GHI} \\
		\hline
		RMSE & RF & \textbf{41.59} & \textbf{88.61} & \textbf{24.17} & \textbf{50.60} & \textbf{77.02} & \textbf{77.55} \\
		& CNN & 45.19 & 89.72 & 30.19 & 51.71 & 78.81 & 79.62 \\
		& LSTM & 47.95 & 95.76 & 26.67 & 50.72 & 77.31 & 78.42 \\
		\hline
		MAE & RF & \textbf{28.17} & \textbf{65.29} & \textbf{16.66} & \textbf{27.52} & \textbf{35.57} & \textbf{41.36} \\
		& CNN & 31.71 & 67.49 & 23.03 & 31.82 & 42.42 & 47.76 \\
		& LSTM & 32.80 & 71.23 & 19.56 & 29.31 & 39.77 & 45.02 \\
		\hline
		nRMSE & RF & \textbf{0.22} & 0.19 & \textbf{0.04} & 0.22 & 0.33 & \textbf{0.19} \\
		& CNN & 0.24 & 0.19 & 0.06 & 0.22 & 0.34 & 0.20 \\
		& LSTM & 0.26 & 0.20 & 0.05 & 0.22 & 0.33 & 0.20 \\
		\hline
		MASE & RF & \textbf{0.23} & \textbf{0.22} & \textbf{0.05} & \textbf{0.15} & \textbf{0.13} & \textbf{0.13} \\
		& CNN & 0.26 & 0.23 & 0.07 & 0.17 & 0.16 & 0.15 \\
		& LSTM & 0.27 & 0.24 & 0.06 & 0.17 & 0.14 & 0.14 \\
		\hline
		% R-squared & RF & \textbf{0.86} & 0.88 & 0.99 & 0.89 & 0.91 & \textbf{0.93} \\
		% & CNN & 0.84 & 0.88 & 0.99 & 0.89 & 0.90 & 0.92 \\
		% & LSTM & 0.82 & 0.86 & 0.99 & 0.89 & 0.91 & 0.92 \\
		% \hline
	\end{tabular}
	\label{tab4}
\end{table}

The evaluation results for the wet season (Table \ref{tab5}) indicate that the LSTM model performed best regarding RMSE, achieving 60.68 for DHI, 91.14 for DNI, and 95.92 for GHI. In contrast, Random Forest (RF) performed better in MAE, with 39.03 for DHI, 47.88 for DNI, and 61.41 for GHI. LSTM and RF tied with nRMSE of 0.27 for DHI and 0.50 for DNI, while GHI results were tied among all models. Also, RF  achieved the lowest MASE for DHI (0.23), DNI (0.19) and GHI (0.21) 

\begin{table}[H]
	\centering
	\caption{Model comparison for wet season forecasting.}
	\begin{tabular}{|l|c|c|c|c|c|c|c|}
		\hline
		\textbf{Metric} & \textbf{Model} & \textbf{Clearsky DHI} & \textbf{Clearsky DNI} & \textbf{Clearsky GHI} & \textbf{DHI} & \textbf{DNI} & \textbf{GHI} \\
		\hline
		RMSE & RF & \textbf{32.97} & 75.49 & \textbf{16.61} & 60.75 & 91.80 & 96.18 \\ 
		& CNN & 36.52 & \textbf{75.22} & 28.94 & 61.40 & 92.59 & 97.42 \\
		& LSTM & 37.48 & 76.64 & 20.33 & \textbf{60.68} & \textbf{91.14} & \textbf{95.92} \\ \hline
		MAE & RF & \textbf{23.08} & \textbf{57.04} & \textbf{12.03} & \textbf{39.03} & \textbf{47.88} & \textbf{61.41} \\
		& CNN & 26.23 & 58.07 & 21.96 & 41.47 & 53.17 & 67.12 \\
		& LSTM & 26.30 & 58.31 & 15.19 & 40.33 & 49.54 & 64.42 \\ \hline
		nRMSE & RF & \textbf{0.22} & 0.13 & \textbf{0.03} & 0.27 & 0.50 & 0.27 \\
		& CNN & 0.24 & 0.13 & 0.05 & 0.28 & 0.51 & 0.27 \\
		& LSTM & 0.23 & 0.13 & 0.04 & 0.27 & 0.50 & 0.27 \\ \hline
		MASE & RF & \textbf{0.31} & 0.23 & \textbf{0.03} & \textbf{0.23} & \textbf{0.19} & \textbf{0.21} \\
		& CNN & 0.35 & 0.23 & 0.07 & 0.24 & 0.21 & 0.23 \\
		& LSTM & 0.36 & 0.23 & 0.05 & 0.24 & 0.20 & 0.22 \\
		\hline
		%R² & RF & 0.76 & 0.89 & 0.997 & 0.84 & 0.86 & 0.86 \\
		%& CNN & 0.70 & 0.89 & 0.99 & 0.83 & 0.85 & 0.86 \\
		%& LSTM & 0.69 & 0.89 & 0.996 & 0.84 & 0.86 & 0.86 \\
		%\hline
	\end{tabular}
	\label{tab5}
\end{table}

Finally, for the dry season forecasting (Table \ref{tab6}), the Random Forest (RF) model consistently outperformed the other models across all metrics. The RMSE values for RF were 37.9 for DHI, 63.5 for DNI, and 54.7 for GHI. For MAE, RF achieved 16.14 for DHI, 25.69 for DNI, and 22.75 for GHI. Similarly, RF had the best nRMSE scores: 0.1544 for DHI, 0.22 for DNI, and tied with CNN with an nRMSE value of 0.12 for GHI. Regarding MASE, RF once again was the best for DHI (0.09), DNI (0.09) and GHI (0.07).

\begin{table}[H]
	\centering
	\caption{Model comparison for dry season forecasting.}
	\begin{tabular}{|l|c|c|c|c|c|c|c|}
		\hline
		\textbf{Metric} & \textbf{Model} & \textbf{Clearsky DHI} & \textbf{Clearsky DNI} & \textbf{Clearsky GHI} & \textbf{DHI} & \textbf{DNI} & \textbf{GHI} \\
		\hline
		RMSE & RF & \textbf{46.52} & \textbf{101.47} & \textbf{29.84} & \textbf{37.91} & \textbf{63.55} & \textbf{54.70} \\
		& CNN & 51.17 & 102.08 & 35.66 & 38.56 & 65.79 & 56.25 \\
		& LSTM & 55.82 & 111.21 & 32.32 & 39.30 & 65.40 & 57.85 \\ \hline
		MAE & RF & \textbf{32.24} & \textbf{75.89} & \textbf{21.65} & \textbf{16.14} & \textbf{25.69} & \textbf{22.75} \\
		& CNN & 36.69 & 77.62 & 27.47 & 21.21 & 34.08 & 29.27 \\
		& LSTM & 39.40 & 84.26 & 24.40 & 19.17 & 30.88 & 28.10 \\ \hline
		nRMSE & RF & \textbf{0.21} & \textbf{0.27} & 0.06 & \textbf{0.15} & \textbf{0.22} & 0.12 \\
		& CNN & 0.23 & 0.27 & 0.07 & 0.16 & 0.23 & 0.12 \\
		& LSTM & 0.23 & 0.29 & 0.06 & 0.16 & 0.23 & 0.13 \\ \hline
		MASE & RF & \textbf{0.22} & 0.27 & \textbf{0.06} & \textbf{0.09} & \textbf{0.09} & \textbf{0.07} \\
		& CNN & 0.24 & 0.27 & 0.08 & 0.12 & 0.12 & 0.09 \\
		& LSTM & 0.26 & 0.29 & 0.07 & 0.12 & 0.11 & 0.09 \\
		\hline
		%R² & RF & 0.88 & 0.8291 & 0.9904 & 0.94 & 0.94 & 0.97 \\
		%& CNN & 0.85 & 0.8271 & 0.9863 & 0.94 & 0.93 & 0.96 \\
		%& LSTM & 0.82 & 0.79 & 0.9888 & 0.94 & 0.93 & 0.96 \\
		%\hline
	\end{tabular}
	\label{tab6}
\end{table}

\subsection{Model prediction}
\begin{figure}[H]
	\centering
	\includegraphics[width=0.85\textwidth]{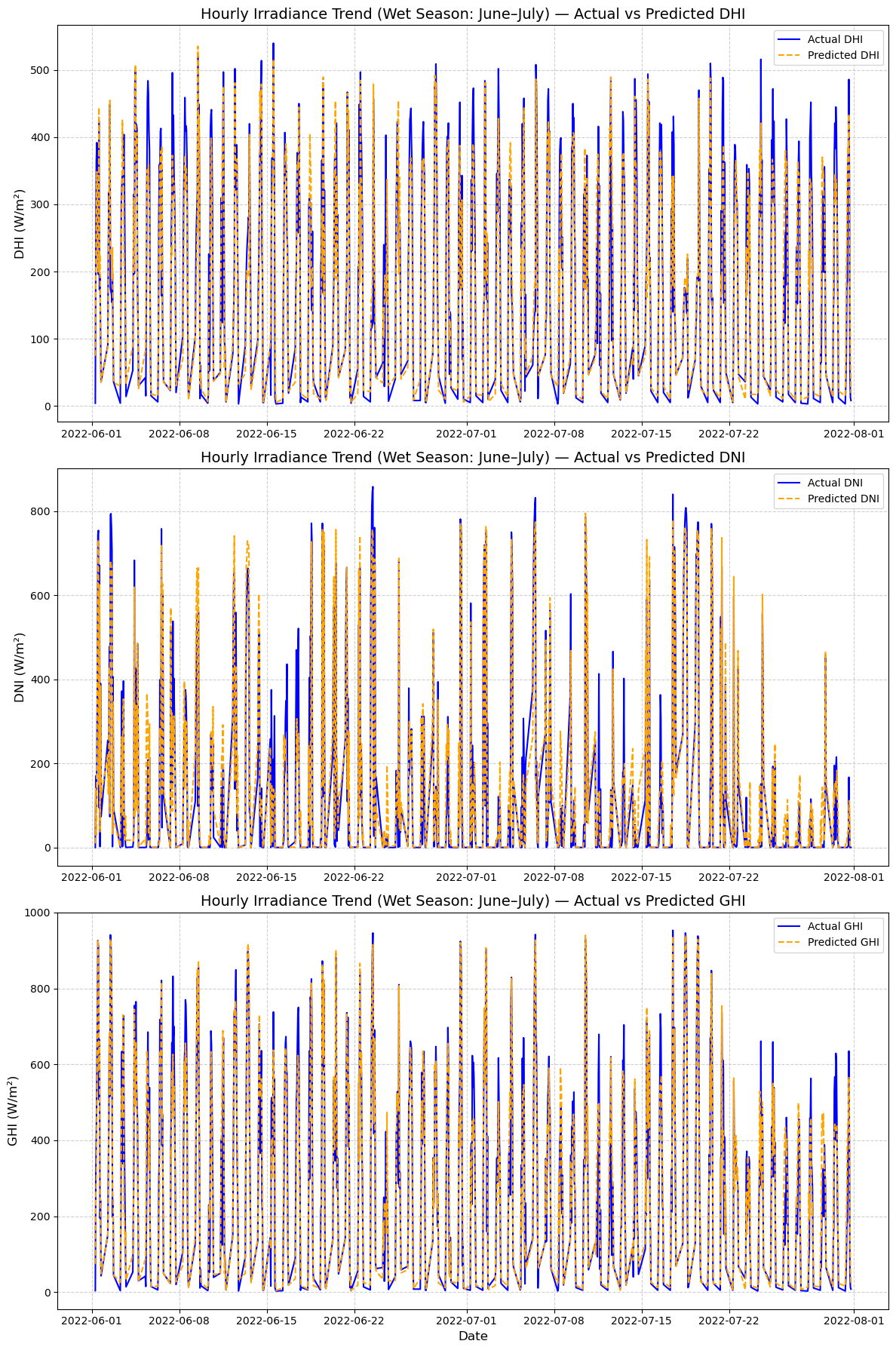}
	\caption{RF annual model prediction for wet season fitted over the actual.}
	\label{fig:rf_prediction_wet_season}
\end{figure}

\begin{figure}[H]
	\centering
	\includegraphics[width=0.85\textwidth]{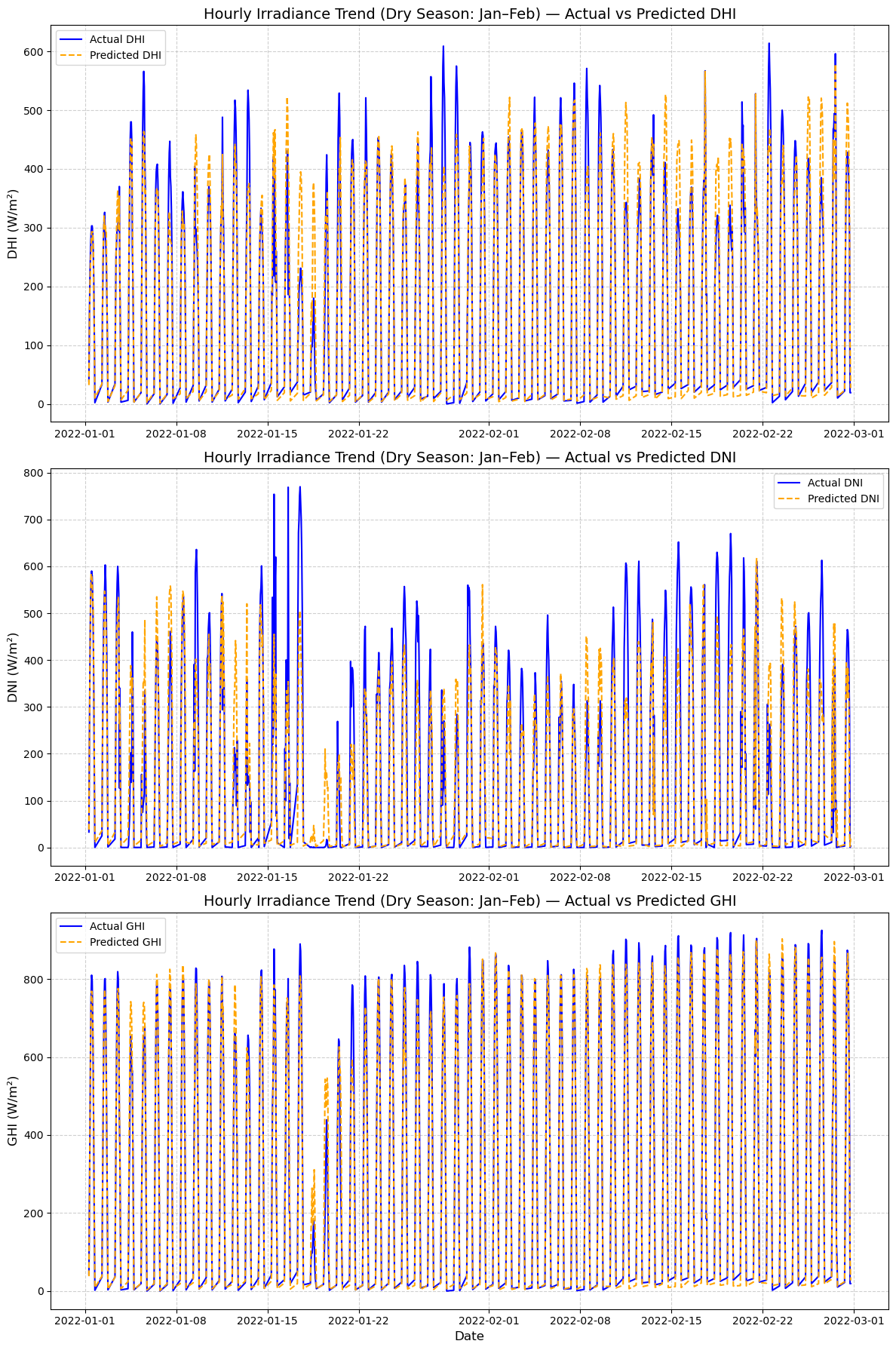}
	\caption{RF annual model prediction for dry season fitted over the actual.}
	\label{fig:rf_prediction_dry_season}
\end{figure}

The Random Forest, the best-performing model, was used to predict the clear-sky radiation components and, subsequently, the actual radiation components.  Figures \ref{fig:rf_prediction_wet_season} and \ref{fig:rf_prediction_dry_season} presents the hourly prediction for wet months (June and July) and dry months (January and February), respectively. While the RF annual model prediction fitted over the actual is given in the appendix in Figure \ref{fig:rf_prediction_full_year}. Similarly Figure \ref{fig:rf_prediction_daily_aggregates} in the appendix illustrates the aggregated daily predictions for DHI, DNI, and GHI, respectively, comparing predicted (orange) and actual (blue) values.

\begin{figure}[H]
	\centering
	\includegraphics[width=0.9\textwidth]{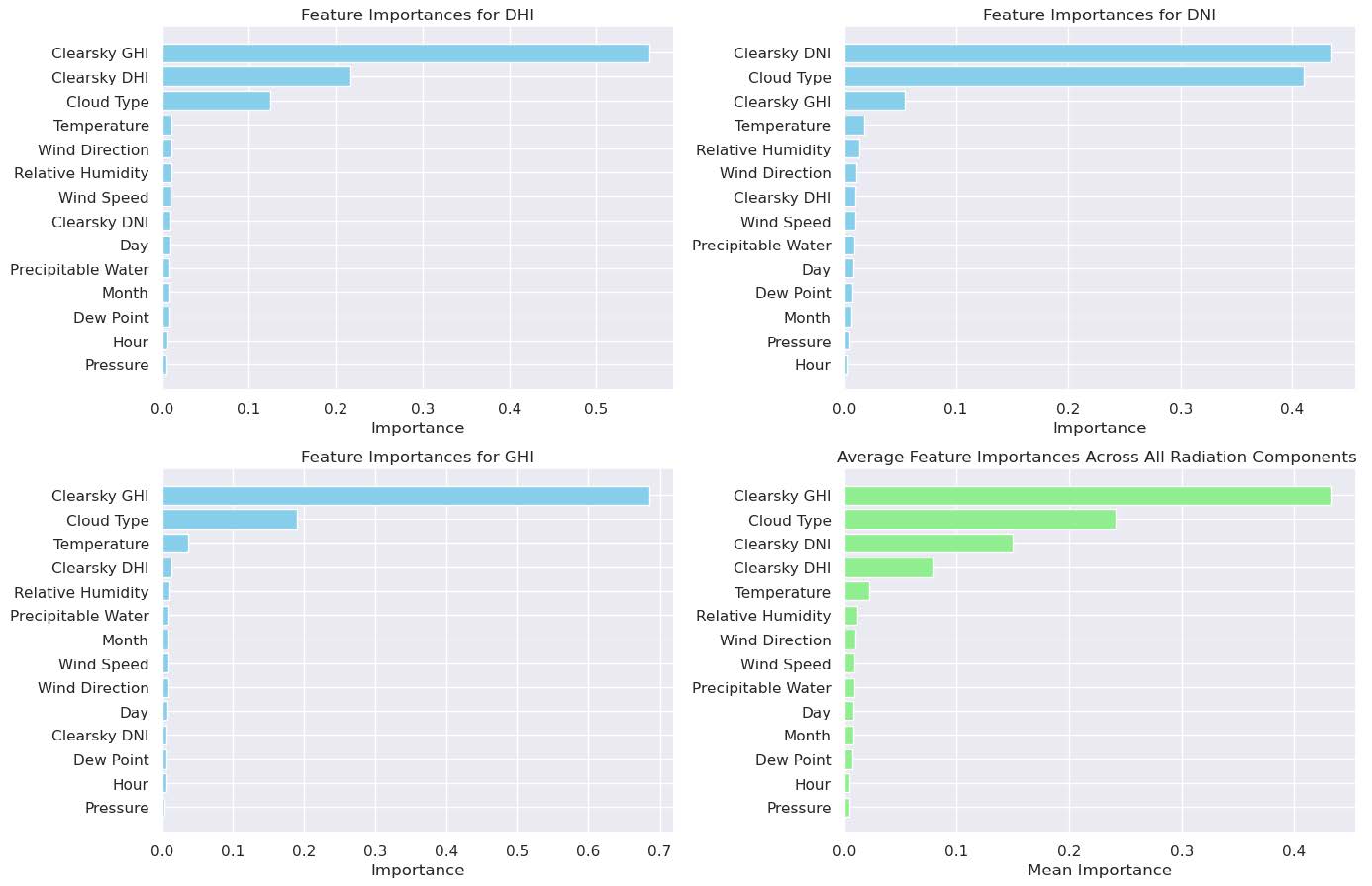}
	\caption{RF feature importance for each and all radiation components.}
	\label{fig:feature_importance}
\end{figure}

Figure \ref{fig:feature_importance} shows the feature importances for DHI, DNI, and GHI and their average, showing that the Clearsky components (DHI, DNI, GHI) are the most significant predictors. Temperature and relative humidity are also important, while wind speed and pressure contribute less. In the correlation analysis, we observed that the Clearsky components correlate with the actual radiation values, though cloud type had weaker correlations (-0.2 for GHI, -0.31 for DNI, and -0.16 for GHI). This suggests that while cloud type does not appear strongly correlated with the radiation components, its importance is amplified in the feature importance plots. Surprisingly, cloud type is more influential than the clear-sky DHI and clear-sky DNI, highlighting its role in solar radiation prediction despite weak correlations.

\subsection{Energy modelling}
\begin{figure}[H]
	\centering
	\includegraphics[width=0.85\textwidth]{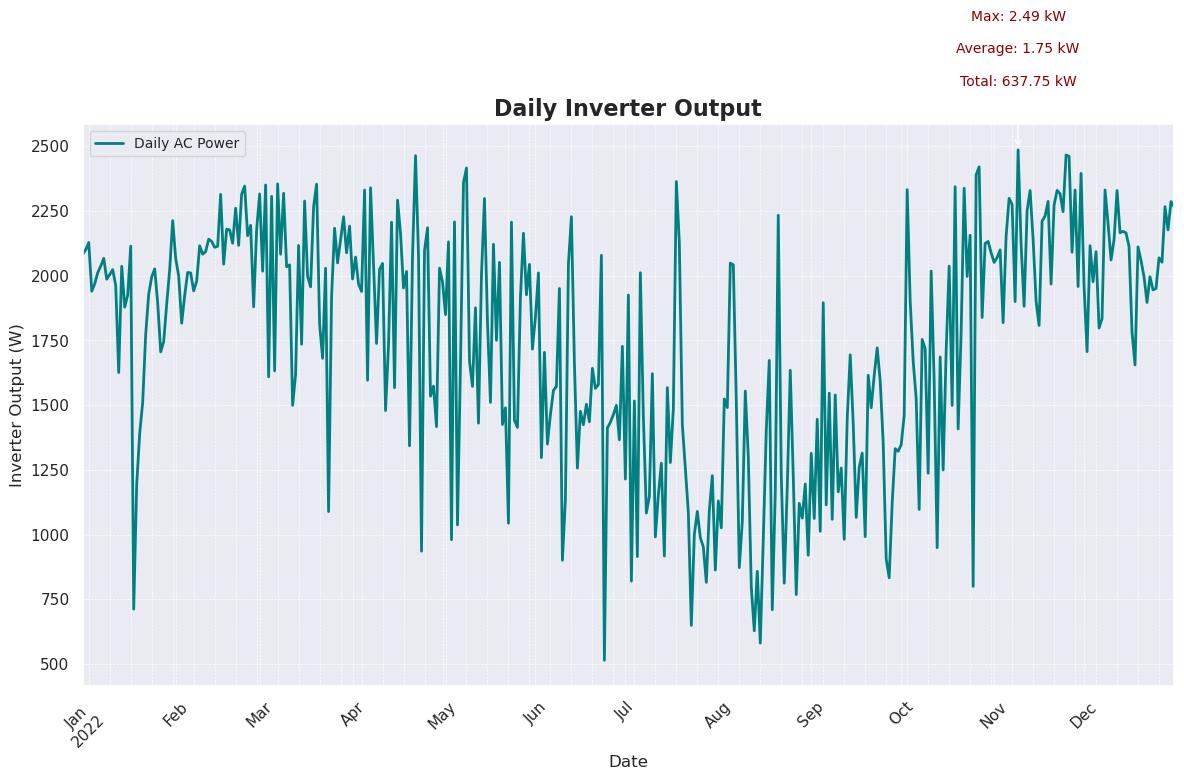}
	\caption{Forecasted daily energy output from TSM-500DE18M (II) module.}
	\label{fig:trina_energy}
\end{figure}

\begin{figure}[H]
	\centering
	\includegraphics[width=0.85\textwidth]{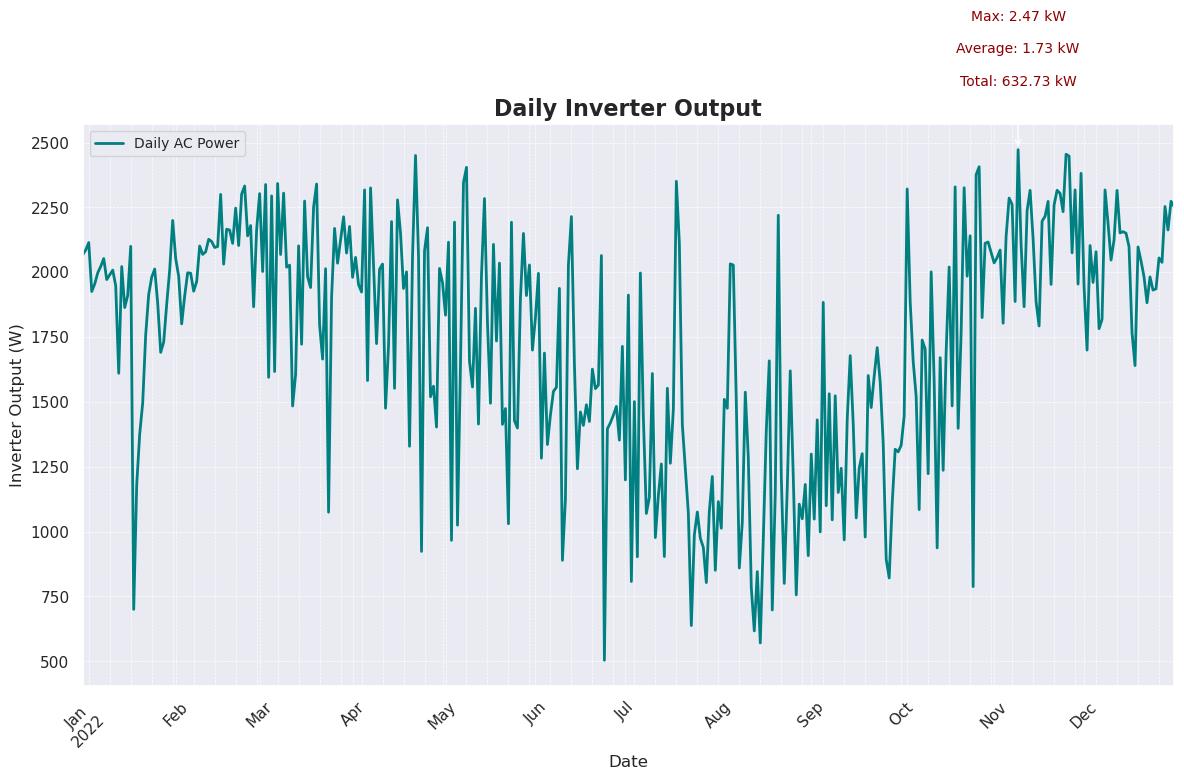}
	\caption{Predicted daily energy output from CS3Y-500MS module.}
	\label{fig:canadian_energy}
\end{figure}

Figures \ref{fig:trina_energy} and \ref{fig:canadian_energy} present the forecasted daily energy output from the Trina Solar and Canadian Solar modules for 2022, derived from the forecasted solar irradiance. The energy outputs for both modules are nearly identical, following the same pattern as the irradiance forecast since energy output is largely dependent on solar radiation.

Using two modules with similar rated power in this study demonstrates the approach’s practicality. It highlights that solar system operators can manually input the system specifications to achieve accurate energy forecasts even when a module is not listed in the CEC database. This flexibility ensures that the methodology can be effectively applied across various solar modules.

\section{Conclusion and future works} \label{sec:4.0}
This work uses three machine learning models—Random Forest, CNN (Convolutional Neural Network), and LSTM—to predict solar radiation. Random Forest is the best performer, achieving an nRMSE of 0.19 for the annual GHI forecasting. These predictions are then integrated with system specifications using PVLIB to model the final energy output. This allows for calculating energy generation ahead of time based on predicted radiation and system parameters, providing a reliable method for energy planning and optimisation.
\par 
Improving the forecasting of Direct Normal Irradiance (DNI), which is still a difficult component of solar radiation forecasting, is one important area for future study. For concentrating solar power technologies, precise DNI prediction is essential to improving the performance and design of solar energy systems. Furthermore, more accurate forecasting is required during the wet season, when increasing cloud cover and rainfall can cause considerable variations in solar energy. The reliability of solar energy systems might be significantly increased by looking into ways to forecast solar radiation more accurately during this time. Furthermore, investigating the use of sophisticated models or hybrid strategies that incorporate several machine-learning techniques may increase the accuracy of forecasts for both seasonal variations and DNI.

\section*{Data and Code Availability Statement}
The data that support the findings of this study are openly available at: \url{https://github.com/peterobarotu/Solar_Energy_Forecasting/tree/main/datasets}.  
In addition, the complete Python source code for data preprocessing, model training, evaluation, and visualization is available at: \url{https://github.com/peterobarotu/Solar_Energy_Forecasting}.

\section*{Authorship Contribution}
\textbf{Obarotu Peter Urhuerhi:} writing - original draft, conceptualization, methodology, data collection, analysis and software.
\textbf{Christopher Udomboso:} writing - methodology, supervision and review.
\textbf{Caston Sigauke:} writing - software, review, editing and final draft.

\section*{Declaration of Competing Interest}
The authors declare that they have no known competing financial interests or personal relationships that could have appeared to influence the work reported in this paper. 

\section*{Acknowledgments}
The authors acknowledge the following for their support:
\begin{enumerate}
	\item[\textbf{i.}] Associates Programme, QLS Section, Abdus Salam International Centre for Theoretical Physics, Trieste, Italy.
	\item[\textbf{ii.}] Centre for Petroleum, Energy Economics and Law, University of Ibadan, Nigeria.
\end{enumerate}

\section*{Abbreviations} 
The following abbreviations are used in this manuscript. \\

\begin{tabular}{@{}ll}
	ARIMA & Autoregressive Integrated Moving Average \\
	MAE & Mean Absolute Error \\
	MASE &  Mean Absolute Scaled Error \\
	RMSE & Root Mean Squared Error \\
	SGB & Stochastic Gradient Boosting \\
	XGBoost & Extreme Gradient Boosting \\
\end{tabular}

\section*{Appendix}

\begin{figure}[H]
	\centering
	\includegraphics[width=0.85\textwidth]{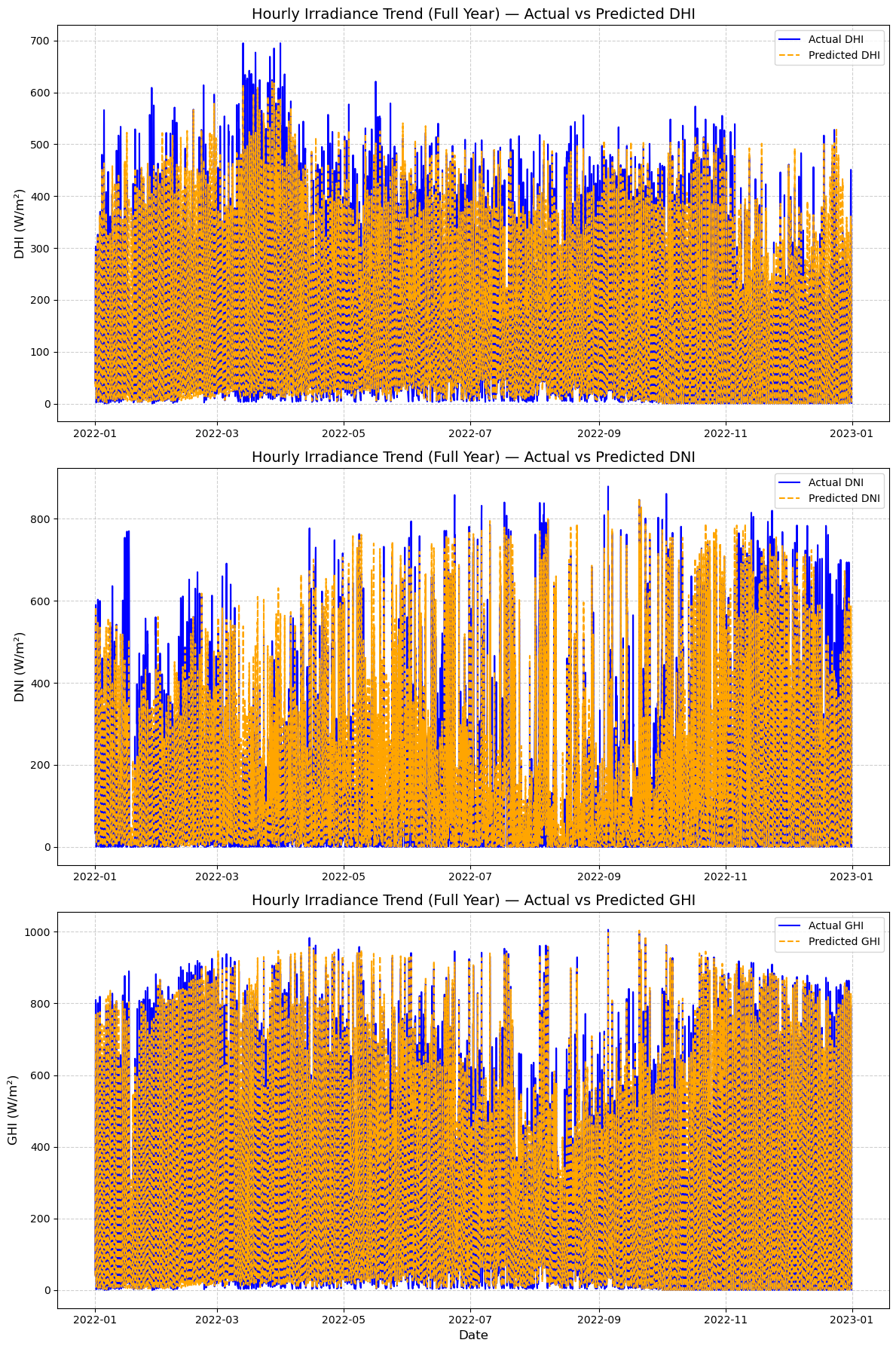}
	\caption{RF annual model prediction fitted over the actual.}
	\label{fig:rf_prediction_full_year}
\end{figure}

\begin{figure}[H]
	\centering
	\includegraphics[width=0.85\textwidth]{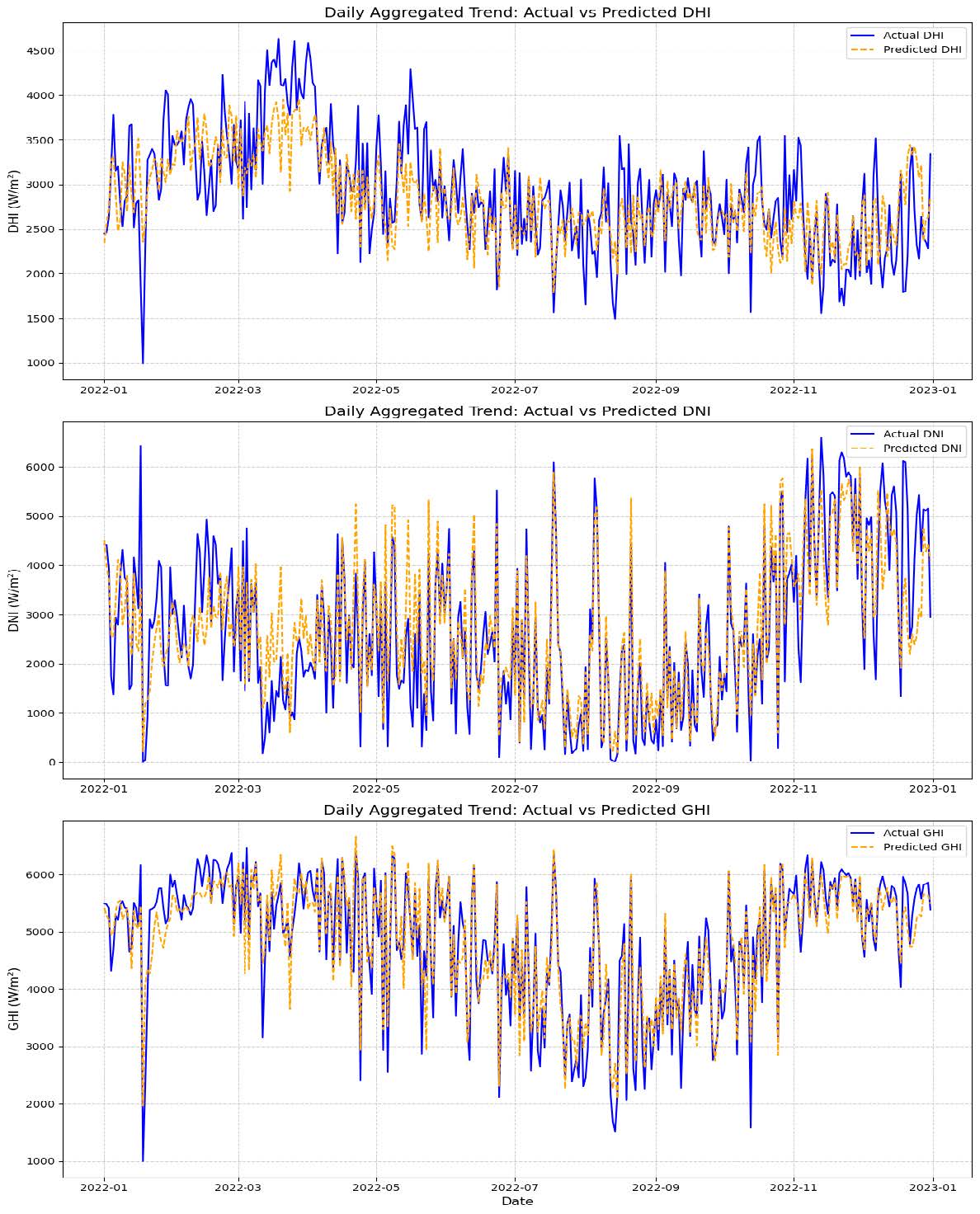}
	\caption{RF annual model prediction (daily aggregated) fitted over the actual.}
	\label{fig:rf_prediction_daily_aggregates}
\end{figure}

\subsection*{Module Specifications: Trina Solar TSM-500DE18M II}
\begin{table}[H]
	\centering
	\renewcommand{\arraystretch}{1.3} % for more vertical spacing
	\caption{Electrical Parameters for Trina Solar TSM-500DE18M II Module}
	\begin{tabularx}{0.75\textwidth}{|l|X|}
		\hline
		\textbf{Parameter} & \textbf{Value} \\
		\hline
		Manufacturer & Trina Solar \\
		Technology & Mono-c-Si \\
		Bifacial & 0 \\
		STC & 500.332 \\
		PTC & 468.3 \\
		A\_c & 2.34 \\
		Length & -- \\
		Width & -- \\
		N\_s & 75 \\
		I\_sc\_ref & 12.28 \\
		V\_oc\_ref & 51.7 \\
		I\_mp\_ref & 11.69 \\
		V\_mp\_ref & 42.8 \\
		alpha\_sc & 0.006754 \\
		beta\_oc & -0.136488 \\
		T\_NOCT & 45.0 \\
		a\_ref & 1.9071 \\
		I\_L\_ref & 12.2823 \\
		I\_o\_ref & 0.0 \\
		R\_s & 0.257757 \\
		R\_sh\_ref & 1373.48 \\
		Adjust & 7.10179 \\
		gamma\_r & -0.337 \\
		BIPV & N \\
		Version & 2023.10.31 \\
		Date & 11/16/2022 \\
		\hline
	\end{tabularx}
	\vspace{0.5em}
	\label{tab:trina_module_parameters}
\end{table}

\subsection*{Module Specification: Canadian Solar CS3Y-500MS}

\begin{table}[H]
	\centering
	\renewcommand{\arraystretch}{1.3} 
	\caption{Electrical Parameters for Canadian Solar CS3Y-500MS Module}
	\begin{tabularx}{0.75\textwidth}{|l|X|}
		\hline
		\textbf{Parameter} & \textbf{Value} \\
		\hline
		Manufacturer & Canadian Solar \\
		Model & CS3Y-500MS \\
		Technology & Mono-c-Si \\
		Bifacial & 0 \\
		STC & 500.4 \\
		PTC & -- \\
		A\_c & -- \\
		Length & -- \\
		Width & -- \\
		N\_s & 78 \\
		I\_sc\_ref & 11.77 \\
		V\_oc\_ref & 53.7 \\
		I\_mp\_ref & 11.12 \\
		V\_mp\_ref & 45.0 \\
		alpha\_sc & 0.005885 \\
		beta\_oc & -0.13962 \\
		T\_NOCT & 45 \\
		gamma\_r & -0.34 \\
		BIPV & N \\
		I\_L\_ref* & 11.77787226112155 \\
		I\_o\_ref* & 2.3707571494843015e-11 \\
		R\_s* & 0.22363339190852521 \\
		R\_sh\_ref* & 334.35946849747984 \\
		a\_ref* & 1.9949684337270626 \\
		Adjust* & 10.480869653415862 \\
		\hline
	\end{tabularx}
	\vspace{0.5em}
	\centerline{\textit{* Parameters generated using PVLib Python.}}
	\label{tab:canadian_parameters}
\end{table}

\subsection*{Inverter Parameters}

\begin{table}[H]
	\centering
	\renewcommand{\arraystretch}{1.3} % for more vertical spacing
	\caption{Electrical Parameters for Fronius Primo GEN24 3.8 208-240 Inverter}
	\begin{tabularx}{0.75\textwidth}{|l|X|}
		\hline
		\textbf{Parameter} & \textbf{Value} \\
		\hline
		Vac & 240 \\
		Pso & 27.8054 \\
		Paco & 3802.0 \\
		Pdco & 3904.29 \\
		Vdco & 400.0 \\
		C0 & -0.000002 \\
		C1 & -0.000033 \\
		C2 & -0.001674 \\
		C3 & -0.000169 \\
		Pnt & 8.3 \\
		Vdcmax & 480.0 \\
		Idcmax & 9.76072 \\
		Mppt\_low & 200.0 \\
		Mppt\_high & 480.0 \\
		CEC\_Date & -- \\
		CEC\_hybrid & Y \\
		\hline
	\end{tabularx}
	\vspace{0.5em}
	\label{tab:fronius_inverter_parameters}
\end{table}

%\renewcommand{\bibname}{References}
%\nocite{*}
%\bibliographystyle{abbrvnat}
%\bibliography{Biblio/References.bib}
%	
%	

\end{document}